\documentclass[preprint,floatfix,aps,showpacs]{revtex4}

\newcommand {\bea}{\begin{eqnarray}}
\newcommand {\eea}{\end{eqnarray}}
\newcommand {\be}{\begin{equation}}
\newcommand {\ee}{\end{equation}}

\usepackage{graphicx}
\usepackage{epsfig,subfigure}
\usepackage{amsmath}
\usepackage{color}
\usepackage{xcolor}
\usepackage[normalem]{ulem}
\usepackage{epstopdf}

\newcommand\redout{\bgroup\markoverwith
{\textcolor{red}{\rule[.5ex]{2pt}{0.4pt}}}\ULon}
\begin{document}
\def\({\left(}
\def\){\right)}
\def\[{\left[}
\def\]{\right]}

\def\Journal#1#2#3#4{{#1} {\bf #2}, #3 (#4)}
\def\RPP{{Rep. Prog. Phys}}
\def\PRC{{Phys. Rev. C}}
\def\PRD{{Phys. Rev. D}}
\def\ZPA{{Z. Phys. A}}
\def\NPA{{Nucl. Phys. A}} 
\def\JPG{{J. Phys. G }}
\def\JPC{{J. Phys. C }}
\def\PRL{{Phys. Rev. Lett}}
\def\PRpt{{Phys. Rep.}}
\def\PLB{{Phys. Lett. B}}
\def\AP{{Ann. Phys (N.Y.)}}
\def\EPJA{{Eur. Phys. J. A}}
\def\NP{{Nucl. Phys}}  
\def\RMP{{Rev. Mod. Phys}}
\def\IJMPE{{Int. J. Mod. Phys. E}}

\input epsf
\title{Impacts of the tensor couplings of $\omega$ and $\rho$  mesons and Coulomb exchange terms on super-heavy nuclei and their relation to symmetry energy }

\author{N. Liliani, A. M. Nugraha, J. P. Diningrum, and A. Sulaksono}

\affiliation{Departemen Fisika, FMIPA, Universitas Indonesia,
Depok, 16424, Indonesia }

\begin{abstract}
We have studied the effects of tensor coupling  of $\omega$ and $\rho$  meson terms, Coulomb exchange term in local density approximation  and  various isoscalar-isovector coupling terms of relativistic mean field model on the properties of nuclear matter, finite nuclei, and super-heavy nuclei. We found that for the same fixed value of symmetry energy $J$ or its slope $L$ the  presence of  tensor coupling  of $\omega$ and $\rho$  meson terms and Coulomb exchange term yields  thicker neutron skin thickness of $^{208}$Pb. We also found that the roles of  tensor coupling  of $\omega$ and $\rho$  meson terms, Coulomb exchange term in local density approximation  and various isoscalar-isovector coupling terms on the bulk properties of finite nuclei varies depending on the corresponding nucleus mass. However, on average, tensor coupling terms play a significant role in predicting the bulk properties of finite nuclei in a quite wide mass range especially in binding energies. We also observed that for some particular nuclei, the  corresponding experimental data of binding energies is rather less compatible with the presence of Coulomb exchange term in local density approximation and  they tend to disfavor the presence of isoscalar-isovector coupling term with too high $\Lambda$ value. Furthermore, we have found that these terms influence the detail properties of $^{292}$120 super-heavy nucleus such as binding energies, the magnitude of two nucleon gaps, single particle spectra, neutron densities, neutron skin thicknesses and mean square charge radii. However, the shell closure predictions of  $^{208}$Pb and  $^{292}$120 nuclei is not affected  by the presence of these terms.
\end{abstract}

\maketitle
\newpage

\section{INTRODUCTION}

\label{sec_intro}
Relativistic mean field (RMF) models (see reviews~\cite{wal,pg,serot,ring2,Meng2006,Meng2016X})  have been successful in providing appropriate description of nuclear matter,  bulk properties as well as deformation for a wide mass spectrum of stable and exotic nuclei. That is because the RMF model provides a natural mechanism for explaining the spin-orbit splitting of single particle states in covariant manner. This feature is very essential in understanding the shell structure of nuclei.

One of challenging issues in nuclear structure physics is to understand the structure of super-heavy nuclei (SHN). From the experimental side, the picobarn ranges of cross sections for production of these nuclei provide limited structural information. On the other hand, the $\alpha$-decay chains of nuclei synthesized in experiments using a $^{48}$Ca beam with actinide targets are stopped by spontaneous fission before reaching the known region of nuclear chart. The problem of unambiguous identification of new isotopes needs to be solved, and more direct techniques to determine $Z$ and $A$ should be used (see for examples Refs.~\cite{SWCD2014, Rudolph2013, Oganessian2013} and the references therein). From the theoretical side, the shell structure of SHN predictions depends on the interplay between the role of strong nuclear attraction among nucleons and the role of Coulomb repulsion from protons added with the fact that for nuclei with very large nucleon number, the spacing among single particle states are very narrow. Therefore, the calculation results depend strongly on the quantum shall effect treatment of model used(see Ref.\cite{SWCD2014} and the references therein). For instance, the microscopic-macroscopic model predicts SHN with $Z$ = 114 and $N$ = 184 as double shell closure~\cite{MN1994}, Non-relativistic Skryme models predict double shell closures at nuclei with $Z$ = 114 and $N$ = 184~\cite{Bender}, with $Z$ = 120 and $N$ = 172~\cite{Bender,RBBSRMG1997,DBGD2003} as well as with $Z$ = 126 and $N$ = 184~\cite{Bender,RBBSRMG1997,Cwiok1996}, while most of the RMF models predict SHN with $Z$ = 120 and $N$ = 172 as double shell closure~\cite{Bender,RBBSRMG1997,Zhang2005,Jiang2010}. It is worthy noting that in the framework of non-relativistic and relativistic self-consistent models, the shell closure predictions are found to be sensitive to the isospin dependence of spin-orbit interaction and the isoscalar effective mass. The uncertainty of these quantities in small nuclei properties amplifies when involving large masses nuclei, while most RMF models provide good spin-orbit splitting throughout the chart of nuclei~\cite{Bender}. It is also important to note that analysis of quasi-particle spectra in $A$ $\sim$ 250 nuclei with spectroscopic data poses additional constraint by defining 'empirical shift' to the energy of spherical states  for the corresponding effective interaction to describe SHN (see Refs.~\cite{Afan06,Afan05} in details). They found that the $Z$ = 120 and  $N$ = 172 SHN prediction of RMF model is compatible with this constraint. They also found that the appearance of large shell gap in  $Z$ = 120 and  $N$ = 172 SHN is due to central depression in its density. The deformed SHN has  also been discussed (see Ref.~\cite{BH2013} and the references therein). Recent review about SHN and fission barriers can be found in Ref.~\cite{Meng2016Y}.
 
The density dependence of symmetry energy is one of other important issues in nuclear physics due to its crucial implications in nuclear and astrophysics (see e.g. Ref.~\cite{HBKLMOPTW14} for recent review). It is reported recently that the softening of the symmetry energy has also important impacts on the empirical shift of spherical states in SHN, the neutron skin thicknesses of SHN, and the central depression of the density of $^{292}$120~\cite{Jiang2010}.  The author of Ref~\cite{Jiang2010}, in his SHN study, adds the term nonlinear isoscalar-isovector coupling to modify the density dependence of symmetry energy prediction of standard RMF model. We need to note that introduction of this term in the RMF model for nuclear matter and finite nuclei applications was done for the first time by the authors of Ref.\cite{Pieka2}. Furthermore, recent study by confronting 263 parametrization of the widely-used RMF models with some experimental and empirically derived nuclear matter constraints shows the crucial role of the isoscalar-isovector coupling term in providing acceptable nuclear matter predictions~\cite{Dutra2014}. 

As discussed in many nuclear physics textbooks, the tensor force plays important role in nucleon-nucleon interaction in free space. It contributes for explaining at the same time the deuteron and many high precision scattering data. Recently it is known that nuclear tensor force also play a crucial role in finite nuclei, particularly in the shell evolution of nuclei\cite{LSMG2008,Otsuka2005,Colo2007,Sagawa2014}. The general form of the potential of the tensor force in coordinate representation is
\begin{equation}
V_T=f(r)S_{12},
\end{equation}
where $S_{12}=3 (\vec{\sigma_1}\cdot \hat{r})(\vec{\sigma_2}\cdot \hat{r})-\vec{\sigma_1}\cdot \vec{\sigma_2}$. The main origin of nuclear tensor interactions stems from $\pi$-nucleon coupling and the tensor parts part of  $\rho$-nucleon and $\omega$-nucleon  couplings~\cite{Sagawa2014}. For instance, in non-relativistic limit, the potential of tensor coupling of $\omega$ and $\rho$ meson terms in coordinate representation takes following form
\begin{eqnarray}
V_\rho&=&f_\rho^2 m_{\rho} \vec{\tau_1}\cdot \vec{\tau_2}\[-\(\frac{1}{3m_\rho r}+\frac{1}{{m_\rho r}^2}+\frac{1}{{m_\rho r}^3}\)e^{-m_\rho r} S_{12}+\frac{2}{3} \vec{\sigma_1}\cdot \vec{\sigma_2}\(\frac {e^{-m_\rho r}}{m_\rho r}-\frac{4 \pi}{m_\rho^3}\delta(r)\)\],\nonumber\\
V_\omega&=&f_\omega^2 m_{\omega} \[-\(\frac{1}{3m_\omega r}+\frac{1}{{m_\omega r}^2}+\frac{1}{{m_\omega r}^3}\)e^{-m_\omega r} S_{12}+\frac{2}{3} \vec{\sigma_1}\cdot \vec{\sigma_2}\(\frac {e^{-m_\omega r}}{m_\omega r}-\frac{4 \pi}{m_\omega^3}\delta(r)\)\].\nonumber\\
\end{eqnarray}
Please see Ref.~\cite{Sagawa2014} and the references therein for detail discussions about the origin and the roles of nuclear tensor force in finite nuclei. The  tensor couplings of $\omega$ and $\rho$ mesons (hereinafter briefly referred to as tensor couplings) were studied in the framework of RMF model for the first time in Ref.~\cite{Rufa}.  NLVT1~\cite{Bender} is one of known parameter sets of standard RMF models with additional tensor coupling terms. Thereafter there are many works performed concerning the effects of these terms on finite nuclei bulk and single particle spectrum (SPS) properties, for instances see Refs.~\cite{Bender,AMBh05,SKBRM,RCMMX95,FRS98,JZZS05}. On one hand, relatively large  contribution of these couplings could improve SPS prediction. However, a moderate value of the  tensor couplings is more demanded to improve the predictions of finite nuclei bulks properties. In the case of SHN, the tensor couplings change the relative spacing between the single particle states, but do not have  significant  effects on spin-orbit splitting~\cite{Bender}. However, in general for RMF approximation, the contribution of $\rho$-tensor coupling is practically negligible. This is due to the fact that the contribution of exchange terms of $\omega$-tensor and $\rho$-tensor couplings as well as $\pi$ meson  is neglected in RMF approximation~\cite{LSGM2007}.

In the RMF framework $\sigma$, $\omega$, $\rho$ and $\delta$  mesons are known to play the role as a mediator for short range interactions due to their quite heavy masses. In the limit of infinite mass of each meson, it would be possible to transform the leading term of meson-mass expansion of the corresponding Fock term into equivalent Hartree forms using the Fierz-transformation~\cite{LKSOR,Maruhn01,SBMRG03}. However, the finite range nature of nuclear force which is indeed encoded in next and next to next leading orders terms (derivative terms) of the corresponding meson-mass expansion.  In principle, localizing of the exchange (Fock) part of these corresponding later terms through Fierz-transformation can be done. But the results can not fully be mapped into  equivalent Hartree forms. The contribution of these next to next leading orders terms  may induce higher order derivative terms in Dirac equation which may violate the relativistic energy-momentum relation. It means that there are some portions of the contribution of exchange terms due to these mesons that cannot be really absorbed by redefining the coupling constants of the complete existing terms in the Hartree approach. This is the reason why, in some cases such as the appearance of some spurious shell structures, e.g., in nuclei with Z=58 and Z=92 can be avoided only by including exchange contributions of $\sigma$, $\omega$, $\rho$,  $\pi$,  and $\rho$-tensor  of mesons within density dependent relativistic Hartree Fock (DDRHF) model~\cite{LSGM2007}. Similarly, the Coulomb exchange contribution cannot also be absorbed into the existing terms in the Hartree approach.  Proper inclusion of Coulomb exchange terms is indeed important for a specific issue~(see Refs.\cite{Liang09,Bloas2011} and the references therein).  It is worthy also here to note that in Skyrme Hartree Fock (SHF) models, the Coulomb exchange contribution is retained by many of SHF parametrization. Recently, the relativistic local density approximation (LDA) for the Coulomb exchange functional in nuclear systems was investigated~\cite{GLLGM2013}. They have found that the important relativistic effects and the exact Coulomb exchange energies can be reproduced by the relativistic LDA within 5 $\%$ demonstrated for semi magic Ca, Ni, Zr, Sn, and Pb isotopes from proton dripline to neutron dripline. It is also important to note, the authors of Ref.~\cite{NLNLG2013} successfully reproduced the exact Coulomb exchange energies by employing the phenomenological formula even with the relative deviations of less than 1 $\%$ for magic Ca, Ni, Sn, and Pb isotopes. To this end, we need also to point out that within the successful DDRHF model and its extension, the relativistic Hartree-Fock-Bogoliubov model, has treated  exactly the Coulomb exchange term (see Ref.~\cite{LRGM2010} and the references therein). A comprehensive discussion on the crucial aspects contained in Fock terms of  relativistic Hartree Fock model which cannot be describe in the Hartree limit such as the one pion exchange, $\rho$-tensor coupling, non-local mean field effects, etc., can be found in Ref.~\cite{Meng2016Z}. 

In this work, first, we would like to investigate the effects of the  tensor coupling terms, relativistic Coulomb  exchange term in LDA, and various isoscalar-isovector coupling terms  within the RMF framework on the properties of nuclear matter, finite nuclei and SHN predictions. Second, we also would like to investigate the impacts of  introducing the  tensor coupling terms and relativistic Coulomb  exchange term  in LDA on the correlation between the density dependence of symmetry energy and the thickness of neutron skin  in SHN. For this purpose we generate some parameter sets which are parametrized with the same protocols.

We organize our works as follows: section~\ref{sec_rmfemtenexc} presents the formalism of RMF model, parametrization of the model, Section~\ref{sec_nucmatfnuc1} presents the  finite nuclei properties and Section~\ref{sec_nucmatfnuc2} presents nuclear matter properties. The discussions of SHN are presented in Sec.~\ref{sec_randd}. The conclusion is given in Sec.~\ref{sec_conclu}.

\section{FORMALISM AND PARAMETRIZATION}

\label{sec_rmfemtenexc}
In this section we briefly review the formalism of RMF model with additional tensor couplings, Coulomb exchange (presented in  LDA) term as well as various isoscalar-isovector couplings. Detailed  derivation for obtaining finite nuclei properties based on standard RMF models as well as the corresponding basic assumption used for example can be found in Refs.~\cite{Meng2016X}. Here we also discuss briefly the parametrization procedure to obtain the parameter sets. 

\subsection{Model Description}
The Lagrangian density of RMF model used in this work is
\begin{equation}
{\mathcal{L}} = {\mathcal{L}}^{\rm free}_{\rm nucleon} + {\mathcal{L}}^{\rm free}_{\rm meson} + 
    {\mathcal{L}}^{\rm lin} + {\mathcal{L}}^{\rm nonlin}+{\mathcal{L}}^{\rm T} + {\mathcal{L}}_{\rm exc}^{\rm C},   
\label{LagM}
\end{equation}
where the free nucleons part in Eq.~(\ref{LagM}) can be expressed as
\begin{equation}
{\mathcal{L}}^{\rm free}_{\rm nucleon} = \bar{\psi} (i\gamma_{\mu}\partial^{\mu} - m_{ N} ) \psi, 
\end{equation}
with $\psi$ and $m_{ N}$  are field and mass of the nucleons, respectively. The mesons part in  Eq.~(\ref{LagM}) can be expressed as
\begin{eqnarray}
{\mathcal{L}}^{\rm free}_{\rm meson} & = & \frac{1}{2}(\partial_{\mu}\Phi\partial^{\mu}\Phi - m_{\sigma}^{2} \Phi^{2}) 
                      \nonumber \\ &-&  \frac{1}{2}(\frac{1}{2} G_{\mu \nu} G^{\mu \nu} - m_{\omega}^{2} V_{\mu} V^{\mu}) \nonumber \\
                      &  -  & \frac{1}{2}(\frac{1}{2} \vec{B}_{\mu \nu} \cdot \vec{B}^{\mu \nu} - m_{\rho} 
^{2} \vec{R}_{\mu} \cdot \vec{R}^{\mu}) 
                     -  \frac{1}{4} F_{\mu \nu} F^{\mu \nu} ,\nonumber\\ 
\end{eqnarray}
where  $\Phi$, $V^{\mu}$, and $\vec{R}_{\nu}$ are $\sigma$, $\omega$ , and $\rho$ meson fields, respectively. The meson tensor fields $G_{\mu \nu}$, $\vec{B}_{\mu \nu}$ and  $F_{\mu \nu}$ are defined as
\begin{eqnarray}
G_{\mu \nu}&=&\partial_{\mu}V_{\nu}- \partial_{\nu}V_{\mu},\nonumber\\
\vec{B}_{\mu \nu}&=&\partial_{\mu}  \vec{R}_{\nu}-\partial_{\nu}  \vec{R}_{\mu}-2 g_\rho \vec{R}_{\mu} \times \vec{R}_{\nu},\nonumber\\
 F_{\mu \nu}&=&\partial_{\mu}A_{\nu}- \partial_{\nu} A_{\mu}.
\end{eqnarray}
Here $A_{\mu}$ is electromagnetic field, while $m_{\sigma}$, $m_{\omega}$ and $m_{\rho}$ are $\sigma$, $\omega$ and $\rho$ mesons masses.  The interactions part in  Eq.~(\ref{LagM}) can be written as 
\begin{eqnarray}
{\mathcal{L}}^{\rm lin} &=& g_{\sigma} \Phi \bar{\psi} \psi - g_{\omega} V_{\mu} \bar{\psi} \gamma^{\mu} \psi 
                          - g_{\rho} \vec{R}_{\mu} \cdot \bar{\psi}
                          \vec{\tau} \gamma^{\mu} \psi\nonumber\\ &-&
                            e A_{\mu} \bar{\psi} \frac{1+\tau_{3}}{2} \gamma^{\mu} \psi,
\end{eqnarray}
where $g_{\sigma}$, $g_{\omega}$, $g_{\rho}$ and $e$ are $\sigma$, $\omega$, $\rho$ and photon coupling constants.  The nonlinear self interactions in  Lagrangian density can be expressed in the following form~\cite{Pieka2,Pieka3,GLLGM2013} 
\begin{eqnarray}
 {\mathcal{L}}^{\rm nonlin} &=& - \frac{1}{3} b_{2} \Phi^{3}
                   - \frac{1}{4} b_{3} \Phi^{4}+\frac{1}{4} c_{3}
                   {(V_{\mu}  V^{\mu})}^2\nonumber\\
                          &+&g_{\omega}^2 g_{\rho}^2 \Lambda {(V_{\mu}  V^{\mu} )(\vec{R}_{\mu} \cdot \vec{R}^{\mu})}  ,
\end{eqnarray}
here $b_{2}$,  $b_{3}$, $c_{3}$ are standard RMF nonlinear parameters and $\Lambda$ is the parameter of isoscalar-isovector coupling term. The tensor couplings in Lagrangian density can be expressed as~\cite{Rufa}  
\begin{equation}
{\mathcal{L}}^{\rm T} =\frac{i f_{\omega}}{2 m_N}\partial_{\nu} V_{\mu} \bar{\psi} \gamma^{\mu}  \gamma^{\nu} \psi  
+\frac{i f_{\rho}}{4 m_N}\partial_{\nu} \vec{R}_{\mu} \bar{\psi} \vec{\tau}
\gamma^{\mu}  \gamma^{\nu} \psi, 
\end{equation}
where $f_{\omega}$ and $f_{\rho}$ are  isoscalar and isovector tensor coupling constants, respectively. Using the same spirit as in SHF model, only the relativistic local density approximation (RLDA) form of the Coulomb exchange energy density is used which can be expressed as ~\cite{Vosko,GLLGM2013} 
\be
<:{\mathcal{L}}_{\rm exc}^{\rm C}:>\approx \frac{3}{4}e^2{(\frac{3}{\pi})}^{1/3}\rho_p^{4/3}\[1-\frac{1}{3 m_N^2}{(3\pi^2)}^{2/3}\rho_p^{2/3}\]~.
\ee

To investigate the role of tensor couplings, Coulomb exchange term and various isoscalar-isovector couplings in nuclear matter, finite nuclei and SHN properties, seven parameter sets are generated. For  each parameter set, except parameter $\Lambda$ which is set arbitrarily, the parameters are obtained through the parametrization procedure. We used almost the same parametrization protocols as used in Ref.~\cite{SKBRM}.  But here we use a larger set of data for parametrization  than those used in Ref.~\cite{SKBRM}. Here, the parametrization data uses, the binding energies of 31 nuclei, rms radii of 21 nuclei, diffraction radii of 18 nuclei with the surface thickness of 16 nuclei. The experimental data for parametrization are taken from Ref.~\cite{Kluepfel}. Here, we use the parametrization weight for binding energies 0.15 $\%$  while  for diffraction and rms radii 0.5 $\%$  for allowed error criteria but  for surface thicknesses, we use the absolute one for allowed error. We also consider the center of mass correction in calculation. The quality of the parametrization will be discussed in the next section. Table~\ref{tab:parset} shows the  obtained parameters of each of parameter set. Here, we name the parameter sets as follows: P0 denotes the parameter set without  tensor couplings, Coulomb exchange term and isoscalar-isovector couplings, PTX are the parameter sets family with tensor couplings and various  isoscalar-isovector couplings  are included where the value of parameter $\Lambda$ is set equal to 0.0X. PTEX family is similar to PTX family but the Coulomb exchange term now is also included. 

\subsection{Parameter Correlations}

It can be seen from Table~\ref{tab:parset} that  for all parameter sets, the values of each isoscalar parameters such as $g_\sigma$,  $g_\omega$, $b_2$, $b_3$, $c_3$ and  $m_\sigma$ does not deviate too much. Therefore, due to the fact that these isoscalar parameters are the dominant contributors for symmetric nuclear matter and finite nuclei properties predictions, the differences in predictions of each specific observable considered in this work are mainly due to the role of $\Lambda$, $g_\rho$, $f_\omega/g_\omega$, and $f_\rho/g_\rho$ as well as due to the contribution of Coulomb exchange term.  It is also worthy to note that there is a correlations between  $\Lambda$, $g_\rho$, $f_\omega/g_\omega$, and $f_\rho/g_\rho$, if the  Coulomb exchange term is included (PTEX family) but only between $\Lambda$, $g_\rho$, and $f_\omega/g_\omega$ if the  Coulomb exchange term is excluded (PTX family). The presence of parameter correlations in isoscalar and isovector sectors is well known problem that occur when adjusting mean field models~\cite{TJB2004}. The reason for these correlations is not their individual values, but rather combinations of the corresponding parameters have physical significance. However, different to the case of isoscalar sector, in isovector sector, the corresponding ``physical significance'' quantities are not yet known. For instance, the authors of Ref.~\cite{TJB2004} demonstrate the strong correlations of isoscalar parameter $\alpha_s$ and $\alpha_v$ of relativistic mean field point coupling model because the sum $\alpha_s$ + $\alpha_v$ determines the isoscalar part of the nuclear central potential, while their difference $\alpha_s$ - $\alpha_v$ determines the size of the spin-orbit potential. It can also be observed that the sign of $f_\rho/g_\rho$ depends on whether the  Coulomb exchange term is present or not. It means that a small portion of the role of Coulomb exchange term contribution, partly can be absorbed by  $\rho$-tensor coupling  when the exchange term contribution is turned off during parametrization. As the consequence of this mechanism, $f_\rho$ does not correlate with $\Lambda$, $g_\rho$, and $f_\omega/g_\omega$ when the  Coulomb exchange term is excluded. However, because the the contribution of $\rho$-tensor coupling are quite small, the change of the sign of $f_\rho/g_\rho$ due above reason does not influence significantly in the quality  of the properties of finite nuclei prediction.

\begin{table} 
\centering
\caption {Parameters of each parameter set obtained in this work through parametrization procedure. Only the value parameters $\Lambda$ are fixed to arbitrary but reasonable numbers, other parameters are freely varied during parametrization. Parametrization status (PS): f means the parameters are freely fitted while c means  to be set arbitrarily. For Coulomb exchange contribution (CEXC):+ means it is included and it is treated self-consistently while - means it is excluded. }
\label{tab:parset}
\begin{tabular}{c | c | c c c| c c c|c }
\hline\hline ~Parameter~&~P0~  &~PT00~  &~PT40~  &~PT55~  &~PTE00~ &~PTE40~ &~PTE60~&PS \\\hline
$g_\sigma$                & 9.87  &9.77  & 9.77  &9.77  & 9.77  & 9.77 &9.77&f\\
$g_\omega$             & 13.18  &13.03  & 13.04 & 13.04 &  13.04 & 13.04 &13.04&f\\
$g_\rho$                  & 5.13  &4.65  & 6.32  &7.25  &4.32  &5.69  &6.63&f\\
$b_2$ (fm$^{-1}$)      & -7.71  &-7.78  &-7.79   &-7.79  &-7.80  &-7.79  &-7.80&f\\
$b_3$                    & 6.20  & 6.37 & 6.24 & 6.19  &6.16 &6.00 &6.02&f\\
$c_3$                    &169.02   &164.70  &166.79  &168.06   &167.92 &166.91 &168.20&f\\
$m_\sigma$ (MeV)          &480.00   &479.40  &479.52  &479.59   &479.71 &479.73 &479.77&f\\\hline
$f_\omega/g_\omega$        &-  &-0.32  &-0.36   &-0.38   &-0.30 &-0.33&-0.35&f\\
$f_\rho/g_\rho$           &-   &-0.008  & -1.19  &-0.51   &1.76 &0.32 &0.04&f\\\hline
$\Lambda$               &-   &0       &0.04   & 0.055  &0     &0.04 &0.06&c\\\hline
CEXC             &-   &-       &-      &-       &+     &+    &+   & \\\hline
\hline
\end{tabular}\\
\end{table}


\section{ FINITE NUCLEI PROPERTIES}

\label{sec_nucmatfnuc1}
Some basic properties of finite nuclei for wide mass range predicted by the corresponding model are given in this section to show the appropriateness of the parameter sets used in this work.

For finite nuclei, besides binding energy $E$, other properties that are considered in this work are rms radius, diffraction radius and surface thickness.
The rms radius is defined as~\cite{Kluepfel}
\begin{equation}
r_{\rm rms}^2=-\frac{3}{F_{\rm ch}}\frac{d^2}{dq^2} F_{\rm ch}(q)|_{q=0},
\end{equation}

\noindent where the $F_{\rm ch}(q)$ is the charge form factor, the (first) diffraction radius
\begin{equation}
R^{\rm diff}=\frac{4.493}{q_0^{(1)}},
\end{equation}
which is determined from the first zero of the charge form factor  $F_{\rm ch}(q_0^{(1)})$=0, and the one of surface thickness is
\begin{equation}
\sigma^2=\frac{2}{q_m} log\[\frac{F_{\rm box}(q_m)}{F_{\rm ch}(q_m)}\], ~ ~ q_m=\frac{5.6}{R},
\end{equation}
here $F_{\rm box}(q)$ corresponds to the form factor of a homogeneous box with radius  $R$.

\subsection{Role of Tensor Coupling Terms, Coulomb Exchange Term and  Various Isoscalar-isovector Coupling Terms}

\begin{figure}
\epsfig{figure=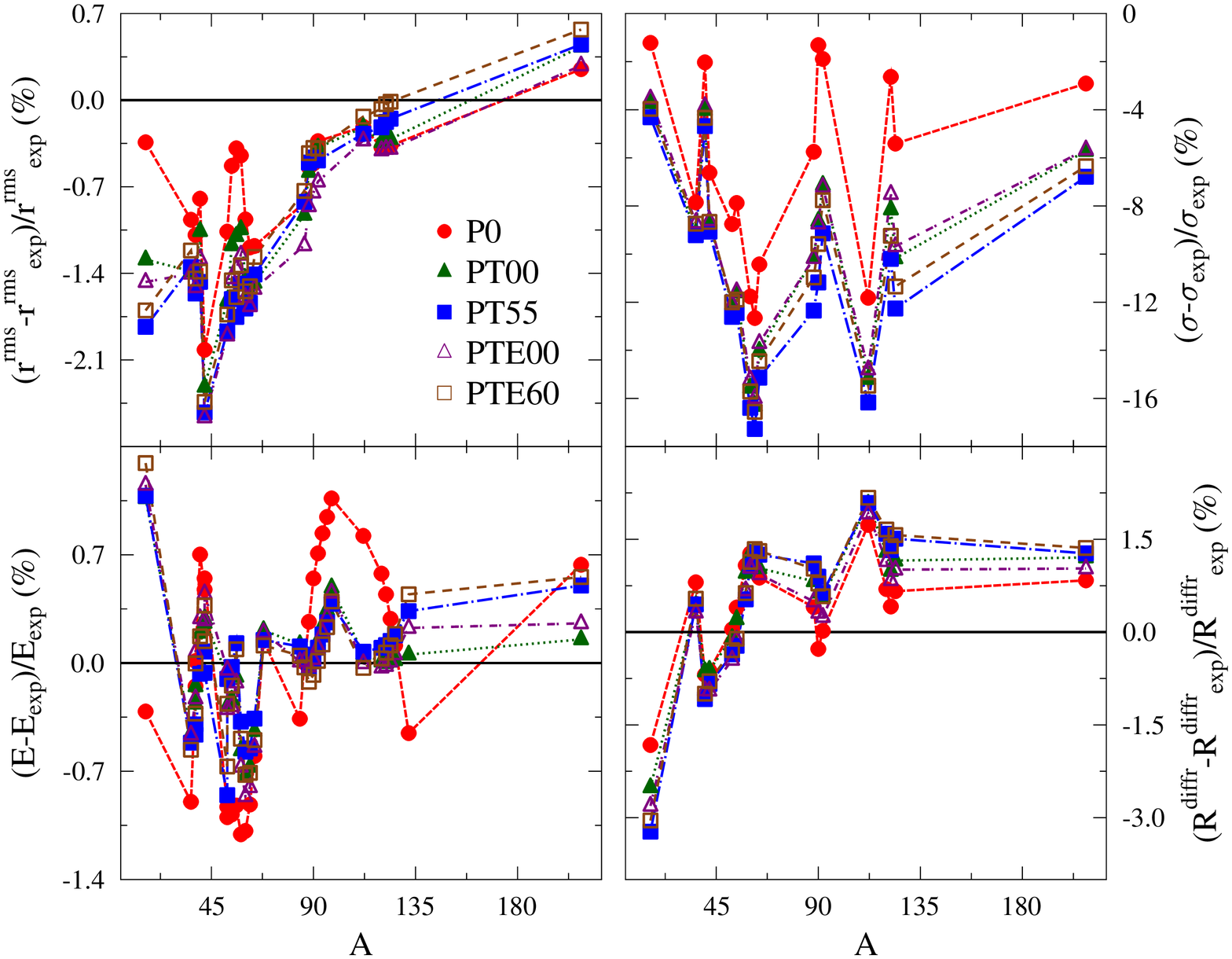, height=15cm, width=16.5cm }
\caption{{\footnotesize(Color online). Global trends of relative error of binding energies (lower left panel), rms radii (upper left panel), surface thicknesses (upper right panel), and diffraction radii (lower right panel) for relatively wide mass range of nuclei predicted by all parameter sets used. }}
\label{Fig:FNSODet}
\end{figure}

For finite nuclei bulk properties such as binding energies, rms radii, diffraction radii, and surface thicknesses, the relative error, i.e., the difference between calculation and experimental values divided by the experimental value in $\%$ can be used as the media to observe the performance or global quality of a parameter set outside its fitting window since this observable is quite sensitive to the differences between used parameter sets predictions.  The compilation of experimental data of binding energies and rms radii were taken from~\cite{Kluepfel} and the references therein. The global trends of relative errors of binding energies,  rms radii, surface thicknesses, and  diffraction radii predicted by each parameter set for relatively wide mass range of nuclei are shown in Fig.~\ref{Fig:FNSODet}. By comparing the results of P0 and those of other parameter sets, it can be seen that the significant role of tensor couplings  for almost  all bulk properties predictions especially for binding energies. Binding energies are the smallest weighted observables in parametrization. For binding energies if tensor couplings are included in parametrization the error is approximately between -1.0 and  0.5 but if these couplings are excluded the error becomes higher i.e., approximately between -1.0 and  1.0. For the diffraction radii and rms radii the effect of tensor couplings does not appear too significant in improving the global quality of prediction. For surface thicknesses, the situation is reversed where the effect of tensor couplings decreases  the quality of the surface thickness prediction. It means that on average, the tensor couplings can relatively improve  the  global quality of binding energy predictions of the model but cannot for other observables, while, the Coulomb exchange term and  isoscalar-isovector couplings contributions on the other hand do not show obvious effects for improving the global quality of the corresponding bulk properties prediction for relatively wide mass range of nuclei. We need to note that all parameter sets are obtained by applying exactly the same parametrization procedure. Therefore, the effects appearing above are genuine due to the corresponding new terms in the model. 

Detailed contribution in each nucleus of tensor couplings, Coulomb exchange term and  isoscalar-isovector coupling might be observed quite clearly from the difference between bulk property predictions with the corresponding effect included and the ones with the corresponding effect excluded. For instances, we can observe the effect of tensor couplings in binding energy from the difference of binding energy predicted by P0 and the one predicted by PT00, while the effect of Coulomb exchange term can be seen from the difference between the binding energy predicted by PT00 and the one by PTE00, and  the effect of  isoscalar-isovector coupling with $\Lambda$=0.0X can be seen from the difference between binding energy predicted by PT00(PTE00) and those of PTX(PTEX). The corresponding results for binding energies, rms radii, diffraction radii, and surface thicknesses are shown in Fig.~\ref{Fig:FNSODetA}. It can be seen clearly that due to self consistent calculation effect and optimization of the role of each term  through parametrization process, the contribution in each bulk property of each corresponding additional term to each nucleus is different. It can be seen on average, for binding energies, diffraction radii, and surface thicknesses, the contribution of tensor couplings is relatively more significant compared to the contribution of other terms, while for the case of isoscalar-isovector couplings, the variation of $\Lambda$ on the corresponding observables appears quite significant only for relatively heavy nuclei. For rms radii, diffraction radii and surface thicknesses, the contribution of  Coulomb exchange term is relatively smaller compared to the contribution of other terms. Interesting to observe however, that on average, the  Coulomb exchange term tends to slightly increase the surface thickness but on the other hand, the tensor couplings and isoscalar-isovector coupling term tend to decrease the surface thickness. Furthermore, it can observe in left panels of Fig.~\ref{Fig:FNSODetA} that the form factor-related observables such as rms radii, diffraction radii and surface thickness are quite sensitive to isoscalar-isovector coupling $\Lambda$ variation. 

\subsection{Interplay among  Tensor Coupling Terms, Coulomb Exchange Term and  Various Isoscalar-isovector Coupling Terms}

\begin{figure}
\epsfig{figure=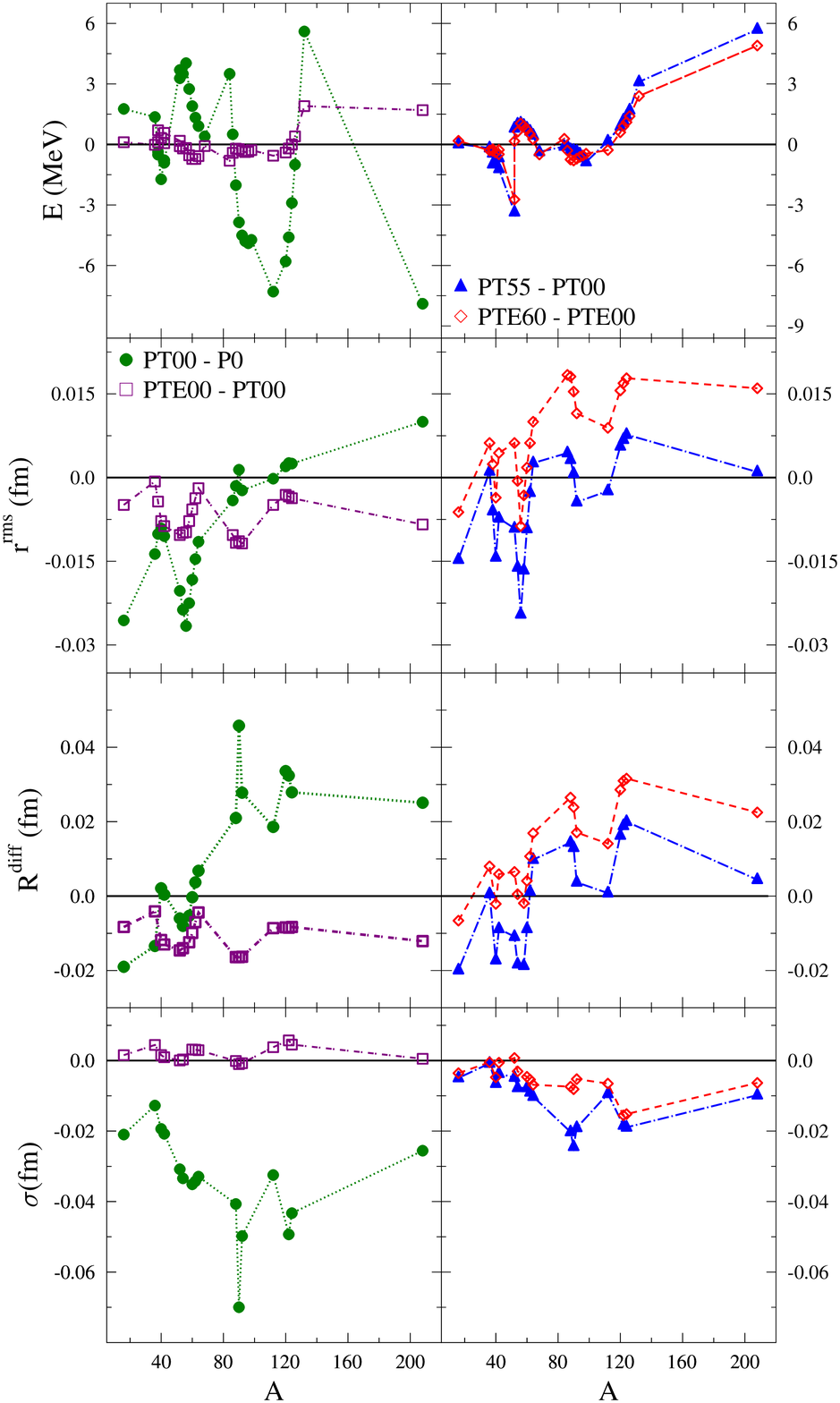, height=20cm, width=14cm }
\caption{{\footnotesize (Color online). Effects of tensor couplings and Coulomb exchange term (left panels) and various isoscalar-isovector couplings (right panels) on relative error of binding energies (upper panels), rms radii and diffraction (middle panels), and surface thicknesses (lower panels), for relatively wide mass range of nuclei.}}
\label{Fig:FNSODetA}
\end{figure}

In general, for every parameter set, the quality of  bulk properties prediction of each nucleus is determined by the interplay of the role of tensor coupling terms, Coulomb exchange term and  various isoscalar-isovector coupling terms. However, the way the of these terms together in yielding finite nuclei bulk property predictions is rather complicated. Because it also depends on which combination of terms are involved as well as it is shown in previous subsection that the contribution due to the role of each term on finite nuclei bulk properties depends on the nucleus mass. For example by comparing the difference between PTE55-PTE00 and PT60-PT00, it can be seen in left panels of Fig.~\ref{Fig:FNSODetA} that when the tensor couplings are included, the interplay between the role of Coulomb exchange term and the role of  isoscalar-isovector couplings provides some differences in rms radii, diffraction radii and surface thicknesses for wide range of A and binding energies with A larger than 120.  

\begin{figure}
\epsfig{figure=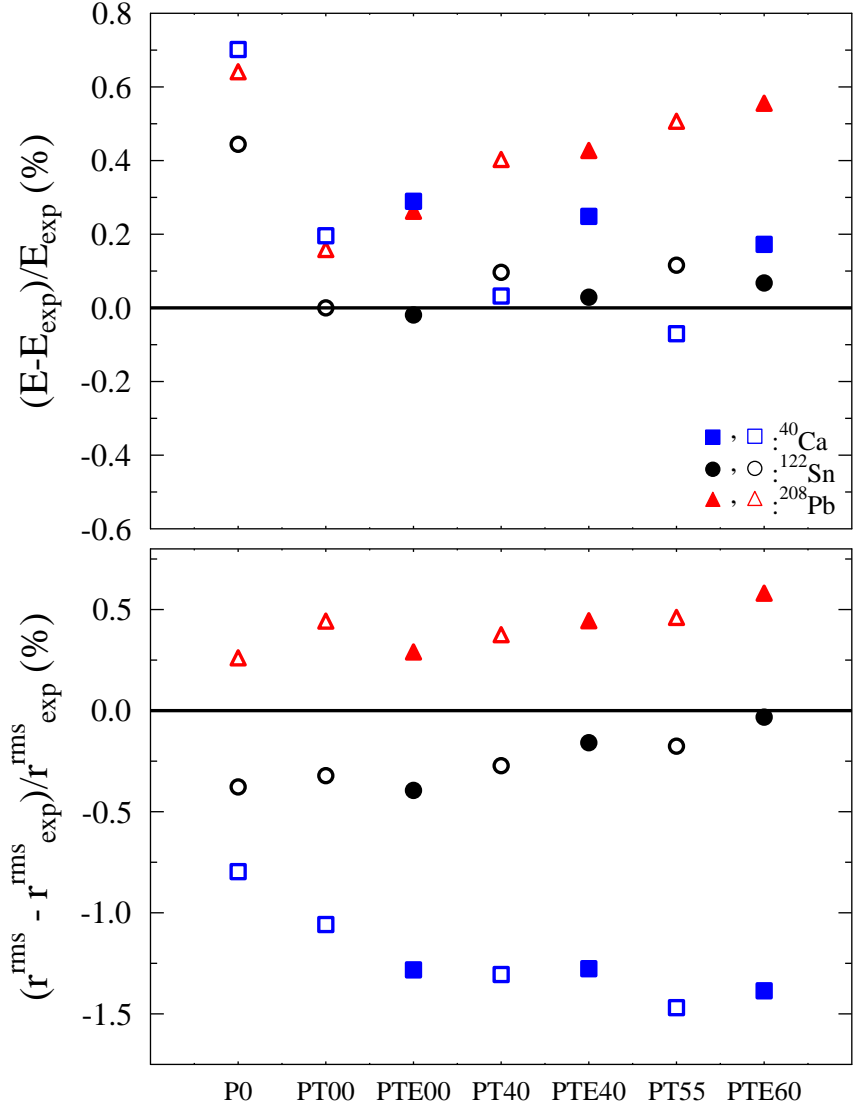, height=17cm, width=12.5cm }
\caption{{\footnotesize (Color online). Effects of tensor couplings, Coulomb exchange term, and isoscalar-isovector coupling parameter variation on relative error of binding energies in lower panel and rms radii in upper panel  of $^{40}$Ca, $^{122}$Sn and  $^{208}$Pb nuclei.}}
\label{Fig:FNSOE}
\end{figure}

To see more details of the interplay role of tensor couplings, Coulomb exchange and isoscalar-isovector coupling terms on the bulk properties of finite nuclei, we present the  relative error of binding energies and rms radii (a representative of form factor-related observables) for $^{40}$Ca as a representative of light nuclei,  $^{122}$Sn as a representative of medium nuclei, and  $^{208}$Pb as a representative of heavy nuclei. The results are shown in Fig.~\ref{Fig:FNSOE}. By comparing the results of P0 and PT00 and PTE00, it is obvious that the role of tensor couplings  significantly decreases  the relative error of binding energies of $^{40}$Ca, $^{122}$Sn and  $^{208}$Pb but  slightly increases  the relative error of  rms radii. For light nuclei like $^{40}$Ca, decreasing the skin thickness by increasing the $\Lambda$ value leads to decreasing its relative error of binding energy, but increasing its relative error of rms radius. Coulomb exchange term makes the binding energy of $^{40}$Ca stronger and this makes  the binding energy different from the prediction and the corresponding experimental data slightly larger. On the other hand,  if we include Coulomb exchange term, the relative error of rms radius prediction is not too sensitive in respect to the $\Lambda$ value variation. For heavy nuclei like $^{208}$Pb, the experimental binding energy is less compatible in the presence of Coulomb exchange term. However, for medium nuclei like $^{122}$Sn, the experimental binding energy indeed favors the presence of Coulomb exchange term. On the other hand, if Coulomb exchange term is included, the rms radius experimental data are consistent with small $\Lambda$ for the  $^{208}$Pb case but relatively large  $\Lambda$ for the  $^{122}$Sn case. However, increasing $\Lambda$ value does not yield visible effect on the rms radius of $^{208}$Pb if Coulomb exchange term is excluded. For $^{132}$Sn and $^{208}$Pb nuclei, increasing $\Lambda$ value leads to increasing relative error of binding energy. The experiment data of binding energies of  $^{132}$Sn and  $^{208}$Pb favor value of $\Lambda$ close to zero but if Coulomb exchange is excluded, the binding energy of light nuclei like $^{40}$Ca favors value of $\Lambda$ close to 0.04. We note that these results show a presence of the competition between the binding energy and the form factor-related observables and trade-off of the quality prediction of each bulk property from light to heavy nuclei. Fitting process optimizes the role of the corresponding terms to accommodate this situation~\cite{TJB2004}.

To this end, it can be concluded that the role of tensor couplings, Coulomb exchange term and isoscalar-isovector couplings  on finite nuclei bulk properties is not the same in every nucleus but they are varied depending on their masses due to self consistent effect in RMF calculation and parameters optimization  through fitting process. In addition, tensor couplings play a significant role in binding energies and surface thicknesses, while apparently in some particular nuclei, experimental data of binding energies are rather less compatible with the presence of Coulomb exchange term in LDA and they tend to disfavor the isoscalar-isovector coupling term with too high $\Lambda$ value.

\section{NUCLEAR MATTER PROPERTIES}
\label{sec_nucmatfnuc2}
Some basic properties of nuclear matter predicted by the corresponding model are given in this section. 

The most precisely determined of symmetric nuclear matter  property is binding energy at saturation density ($E$). Other nuclear matter isoscalar properties at saturation  density can be derived from binding energy $E (\rho)$ because they are defined as 
\bea
K_0 &=& 9 \rho_0^2 \frac{d^2 E(\rho)}{d\rho^2} |_{\rho=\rho_0},\nonumber\\
J_0  &=& 27 \rho_0^3 \frac{d^3 E(\rho)}{d\rho^3} |_{\rho=\rho_0},
\label{Eq:NMISCR}
\eea 
while  in isovector sector, the role of symmetry energy  at the saturation density of $J$ is very crucial. Other nuclear matter isovector properties at saturation density  can be derived from $J (\rho)$ which are obtained from the following relations
\bea
L &=& 3 \rho_0 \frac{d J (\rho)}{d\rho} |_{\rho=\rho_0},\nonumber\\
K_{\rm sym} &=& 9 \rho_0^2 \frac{d^2 J (\rho)}{d\rho}^2 |_{\rho=\rho_0},\nonumber\\
K_{asy}&=&K_{\rm sym}-6 L ,\nonumber\\
K_{\rm sat,2}&=&K_{asy}-\frac{J_0}{K_0}L.
\label{Eq:NMIVEC}
\eea 
Here $E$ and $\rho_0$ denote nuclear matter binding energy and saturation density, respectively. Table~\ref{tab:nucmatprop} shows nuclear matter properties at $\rho_0$. It is obvious that all parameter sets predict more or less similar $k_F $, $E$, and $K_0$. However, It can be seen that $J$, $L$, $K_{\rm asy}$ and $K_{\rm sat,2}$ decrease when the value of $\Lambda$ parameter increases.  For a fixed value of $\Lambda$, implicit inclusion of tensor couplings and Coulomb exchange term  also tends to slightly decrease the  $K_0$, $J$, $L$, $K_{\rm asy}$ and $K_{\rm sat,2}$ value, respectively.  
\begin{table}
\centering
\caption {Nuclear matter properties at saturation density ( $\rho_0$): Fermi momentum ($k_F$), binding energy ($E$), incompressibility coefficient for symmetric nuclear matter ($K_0$), symmetry energy ($J$) and other quantities defined by Eqs. (\ref{Eq:NMISCR})-(\ref{Eq:NMIVEC}). }
\label{tab:nucmatprop}
\begin{tabular}{c | c | c c c | c c c }
\hline\hline ~Parameter~&~P0~  &~PT00~  &~PT40~  &~PT55~  &~PTE00~ &~PTE04~ &~PTE06~ \\\hline
$k_F (fm^{-1})$                &1.3   &1.3  &1.3  &1.3  &1.3  &1.3 &1.3\\
$E$ (MeV)                &-16.6   &-16.2  &-16.2  &-16.1  &-15.9  &-15.9 &-15.8\\
$K_0$ (MeV)               &231.2   &220.0  &220.0  &223.7  &219.0  &220.0 &218.0\\
$J$ (MeV)                &43.4   & 37.8 &31.3  &29.6  &35.4  &30.2 &28.1\\
$L$ (MeV)               & 130.0  & 112.7 &49.0  &42.3  & 105.4 &52.3 &43.6\\
$K_{\rm asy}$  (MeV)              &-754.3   &-652.8  & -309.0 &-213.4  &-608.5  &-340.7 &-230.8\\
$K_{\rm sat,2}$ (MeV)               &-528.9   &-437.9  &-216.8  & -130.8 & -412.0 &-243.7 &-148.4\\\hline\hline
\end{tabular}\\
\end{table}

\begin{figure}
\epsfig{figure=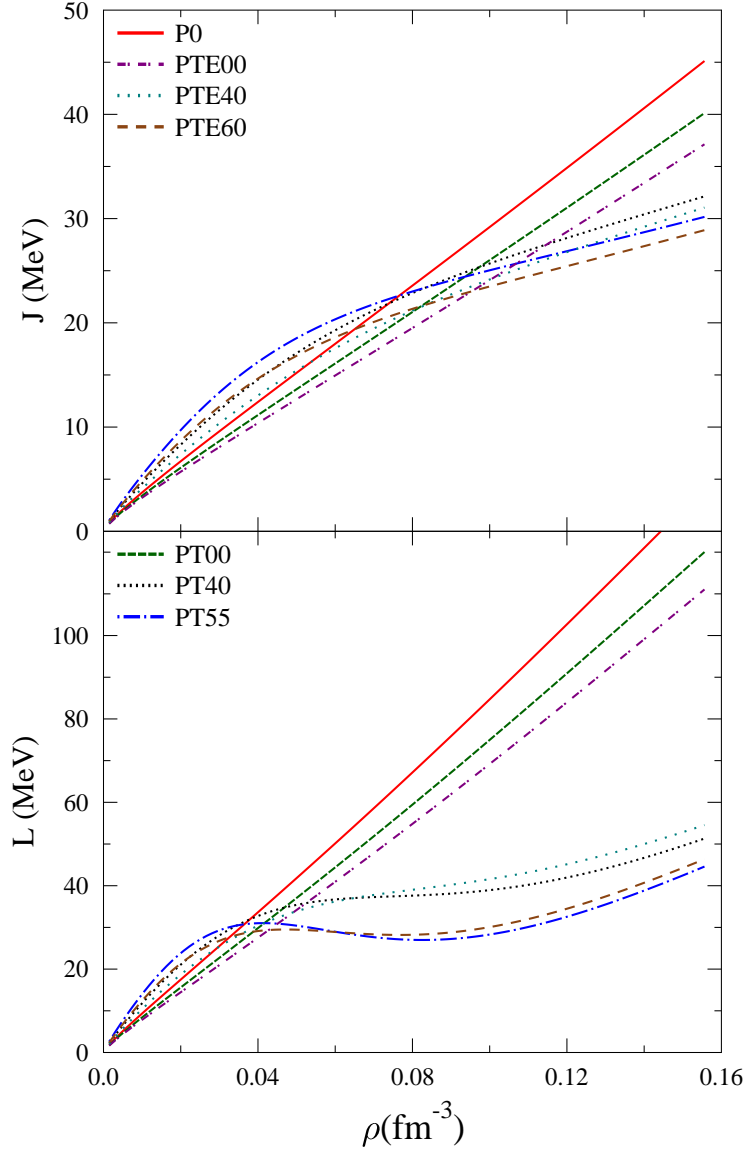, height=16cm, width=10cm }
\caption{{\footnotesize (Color online). Effects of tensor couplings and Coulomb exchange term on the density dependence of symmetry energy ($J(\rho)$) in upper panel and its slope ($L(\rho)$) with various isoscalar-isovector couplings are shown  in lower panel. }}
\label{Fig:NMsymE}
\end{figure}

\begin{figure}
\epsfig{figure=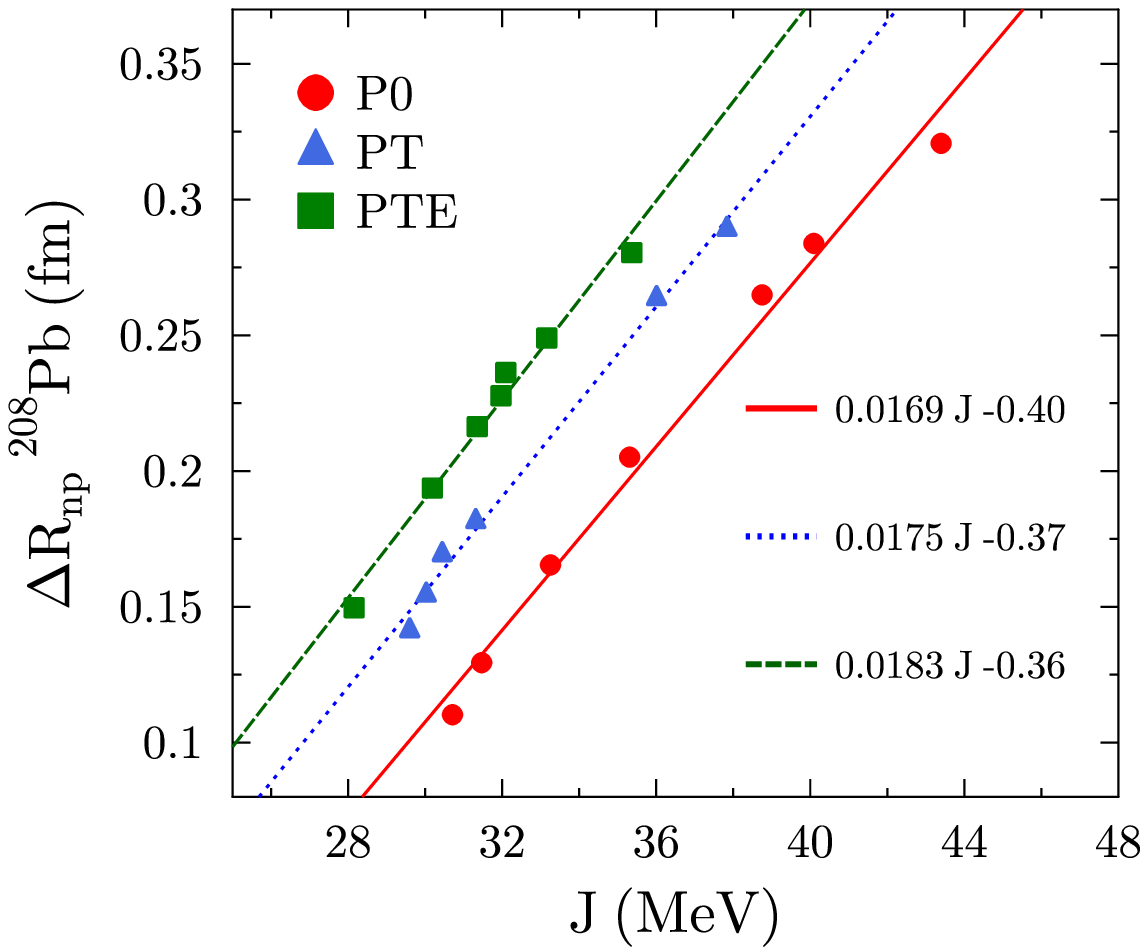, height=11.5cm, width=13cm }
\caption{{\footnotesize(Color online). Effects of tensor couplings and Coulomb exchange term with various isoscalar-isovector couplings on the correlation between the neutron skin thicknesses of $^{208}$Pb and symmetry energy value at saturation. P0 denotes parameter set with tensor couplings and Coulomb exchange term are excluded, PT denotes parameter set with tensor term is included and PTE denotes parameter set with tensor couplings and Coulomb exchange term are included.}}
\label{Fig:SkinsymC}
\end{figure}

The density dependence of $J(\rho)$ and $L(\rho)$ predicted by used parameter sets are shown in upper and lower panels of Fig.~\ref{Fig:NMsymE}, respectively. The role of isoscalar-isovector coupling term to make the relationships between $J(\rho)$ or $L(\rho)$ and $\rho$ nonlinear is obvious. As a consequence, the $J(\rho)$ and $L(\rho)$ become softer if isoscalar-isovector coupling term is included. If the isoscalar-isovector coupling term is excluded as in the cases of P0, PT00 and PTE00 parameter sets, the roles of  tensor couplings and Coulomb exchange term appear clear i.e.,  they make the slope of $J(\rho)$ or $L(\rho)$ and $\rho$ relations change. The implication of the presence of  tensor couplings and Coulomb exchange term on the correlations between the neutron skin thickness of $^{208}$Pb and symmetry energy at $\rho_0$  is shown in Fig.\ref{Fig:SkinsymC}. The slope of the neutron skin thickness of $^{208}$Pb and $J$ linear relation do not change much by the presence of tensor couplings and Coulomb exchange term during parametrization but  they change the  value of constant of the  corresponding linear relation. It means that for the same fixed value of $J$ or $L$, the presence of  tensors couplings and/or Coulomb exchange term makes the thickness of neutron skin  of $^{208}$Pb predictions thicker. 


\section{EXTRAPOLATION TO SUPER-HEAVY NUCLEI}
\label{sec_randd}
In this section, we study how the tensor couplings, various isoscalar-isovector couplings and Coulomb exchange term affect the prediction of the double magic SHN. We  check the existence of shell closures of SHN by observing the existence of peak in two nucleons gaps and the existence of large gaps in closed shells near Fermi energies of the corresponding nuclei. In addition, we also show the  profiles of proton and neutron densities as well as the surface thickness of the corresponding nuclei. The corresponding properties of well-known double magic heavy nucleus i.e., $^{208}$Pb are also calculated for standard reference. 

\subsection{Two Nucleon Gaps}
The expressions for two neutrons gap can be written as  
\begin{eqnarray}
\delta_{2n}=2E(N,Z)-E(N-2,Z)-E(N+2,Z),
\end{eqnarray}
while the one for  two protons gap can be written as
\begin{eqnarray}
\delta_{2p}=2E(N,Z)-E(N,Z-2)-E(N,Z+2).
\end{eqnarray}

\begin{figure}
\epsfig{figure=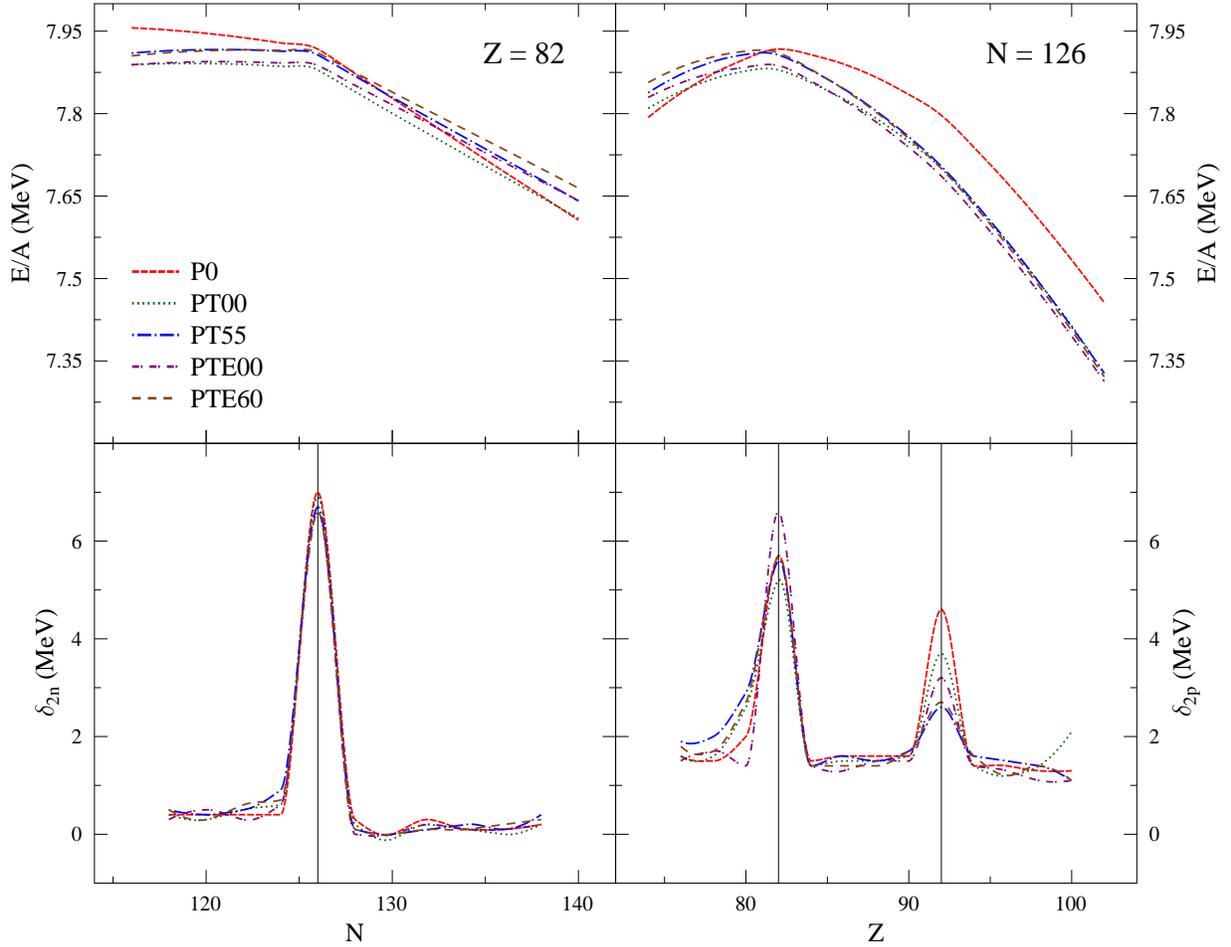, height=13.5cm, width=16.5cm }
\caption{{\footnotesize (Color online). Binding energies, two-neutron and two-proton gaps of the chain of $Z$ = 82 isotopes (left panels) and $N$ = 126 isotones (right panels) predicted by all used parameter sets. }}
\label{Fig:SHPb}
\end{figure}

\begin{figure}
\epsfig{figure=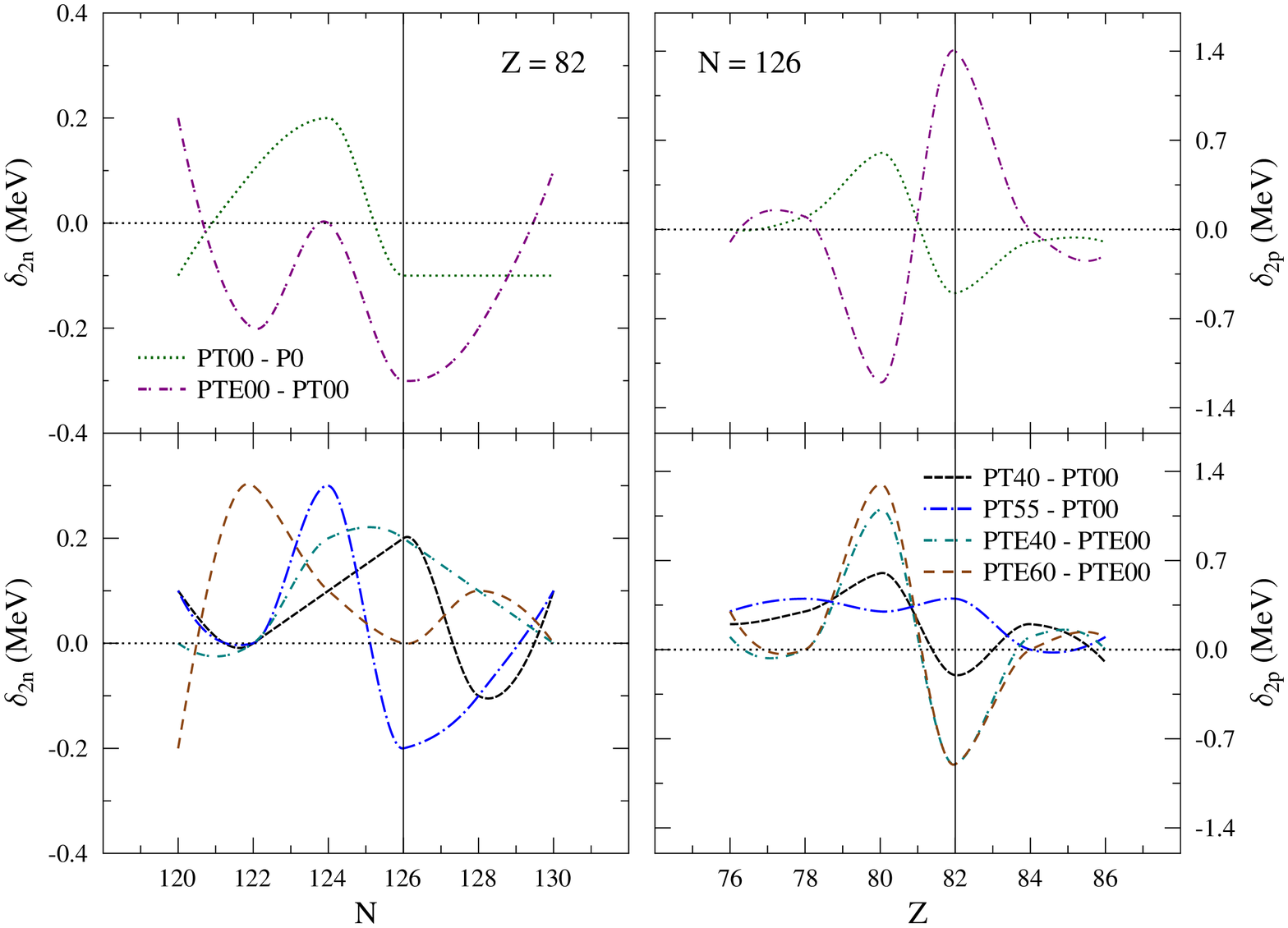, height=13.0cm, width=16.5cm }
\caption{{\footnotesize (Color online). Effect of tensor couplings and Coulomb exchange term (upper panels), and various isoscalar-isovector couplings (lower panels) on two-neutron (left panel) and two proton (right panel) gaps of $Z$ = 82 isotopes and $N$ = 126 isotones. }}
\label{Fig:SHPb1}
\end{figure}

Fig.~\ref{Fig:SHPb}, shows the binding energies in upper panels and two nucleon gaps in lower panels, respectively, with  the chains of $Z$ = 82 isotopes and  $N$ = 126 isotones. It can be observed that for $Z$ = 82 isotopes, the significant effect do by  tensor couplings to decrease the binding energy in the area $N$ $\le$ 126. While for $N$ = 126 isotones, the effect of tensor couplings appears in $Z$ $\ge$ 82. These  terms only marginally  influence the $\delta_{2n}$ and $\delta_{2p}$ of these chains of isotopes and isotones. It can be observed that along the chains of $Z$ = 82 isotopes on the left lower panel that the peak in $\delta_{2n}$ is predicted by all parameter sets  to appear at $N$ = 126  and  along the chains of $N$ = 126 isotones on the right panel, the largest $\delta_{2p}$ peak is at $Z$ = 82, respectively. Those peaks signal that the $^{208}$Pb  has closed shells\cite{Jiang2010}.  These terms only increase the magnitude of the $\delta_{2p}$ peak at $Z$ = 82 and decrease the magnitude at $Z$ = 92.  Therefore, according to this $\delta_{2p/2n}$  peaks observation, the presence of tensor coupling terms, Coulomb exchange term and isoscalar-isovector coupling terms do not influence the shell closures prediction of $^{208}$Pb. The detailed  role of these terms for nuclei around both peaks can be seen in Fig.~\ref{Fig:SHPb1}.  The effect of tensor couplings can be seen from the difference between the $\delta_{2p/2n}$ obtained by PT00 and those obtained by P0, the effect of Coulomb exchange term can be seen from the difference between the $\delta_{2p/2n}$  obtained by  PTE00  and those  obtained by PT00, while the effect of isoscalar-isovector coupling with $\Lambda$ = 0.0X can be seen from the difference between the   $\delta_{2p/2n}$ obtained by PTX(PTEX)  and those  obtained by PT00(PTE00). It can be observed that for  $Z$ = 82 isotopes, for the case  $\Lambda$ = 0, the parameter sets with tensor couplings included tend to relatively decrease  the magnitude of the N=126 peak. However, on the contrary, for the case  of non zero $\Lambda$, they tend to relatively increase  the magnitude of the peak if $\Lambda$ is smaller. On the other hand, for  $N$ = 126 isotones, for the case  of $\Lambda$ = 0, the role of tensor couplings tends to relatively decrease  but the role of exchange Coulomb term tends to increase relatively the magnitude of the gap. On the other hand, the role of isoscalar-isovector coupling is decreasing relatively the magnitude of the gap. For both isotopes and isotones, for non zero $\Lambda$ cases, the quantitative effect depends on the presence of Coulomb exchange term.

\begin{figure}
\epsfig{figure=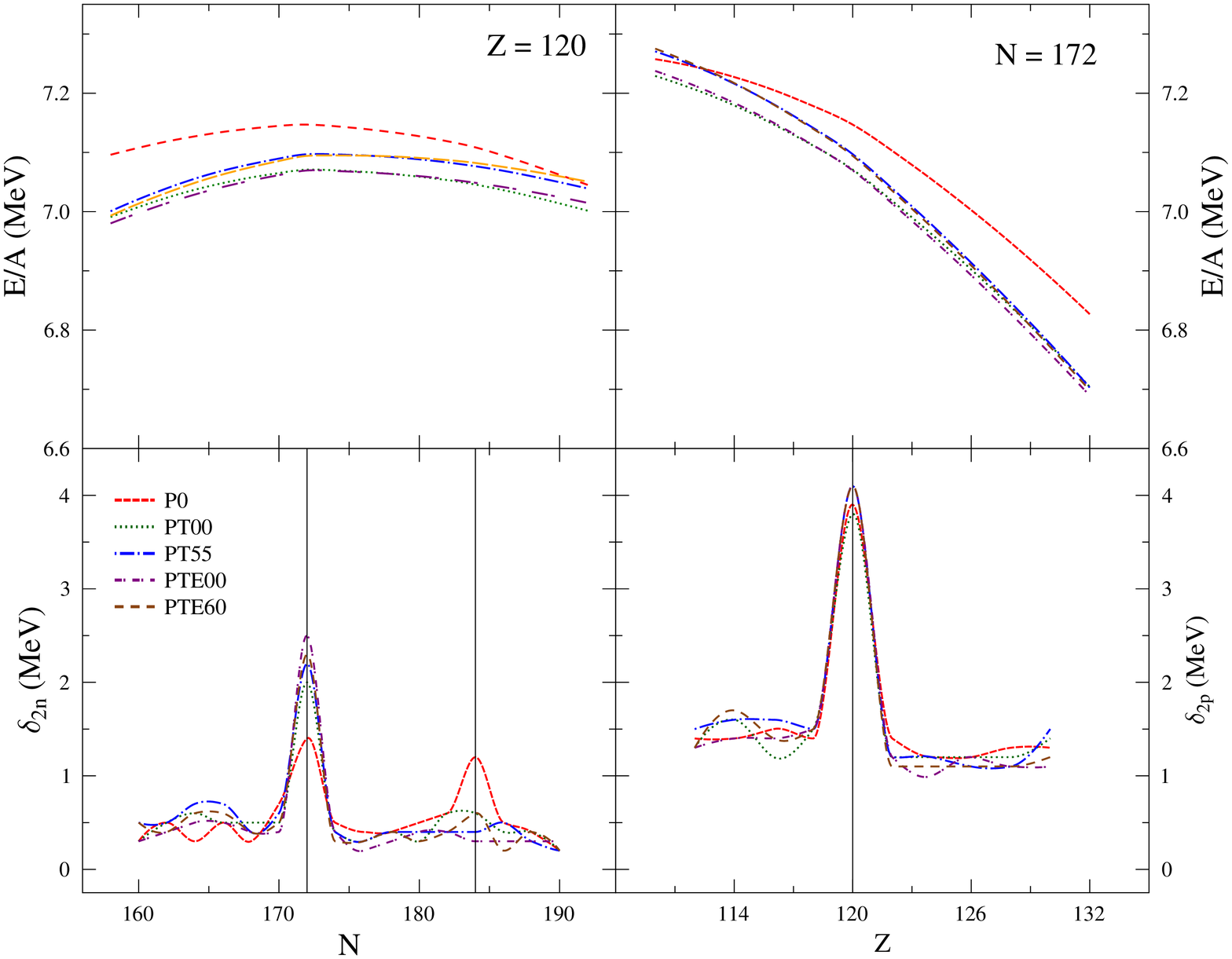, height=13.0cm, width=15.5cm }
\caption{{\footnotesize (Color online). The same as in Fig.~\ref{Fig:SHPb} but for the chain of $Z$ = 120 isotopes and $N$ = 172 isotones.  }}
\label{Fig:SH120}
\end{figure}

\begin{figure}
\epsfig{figure=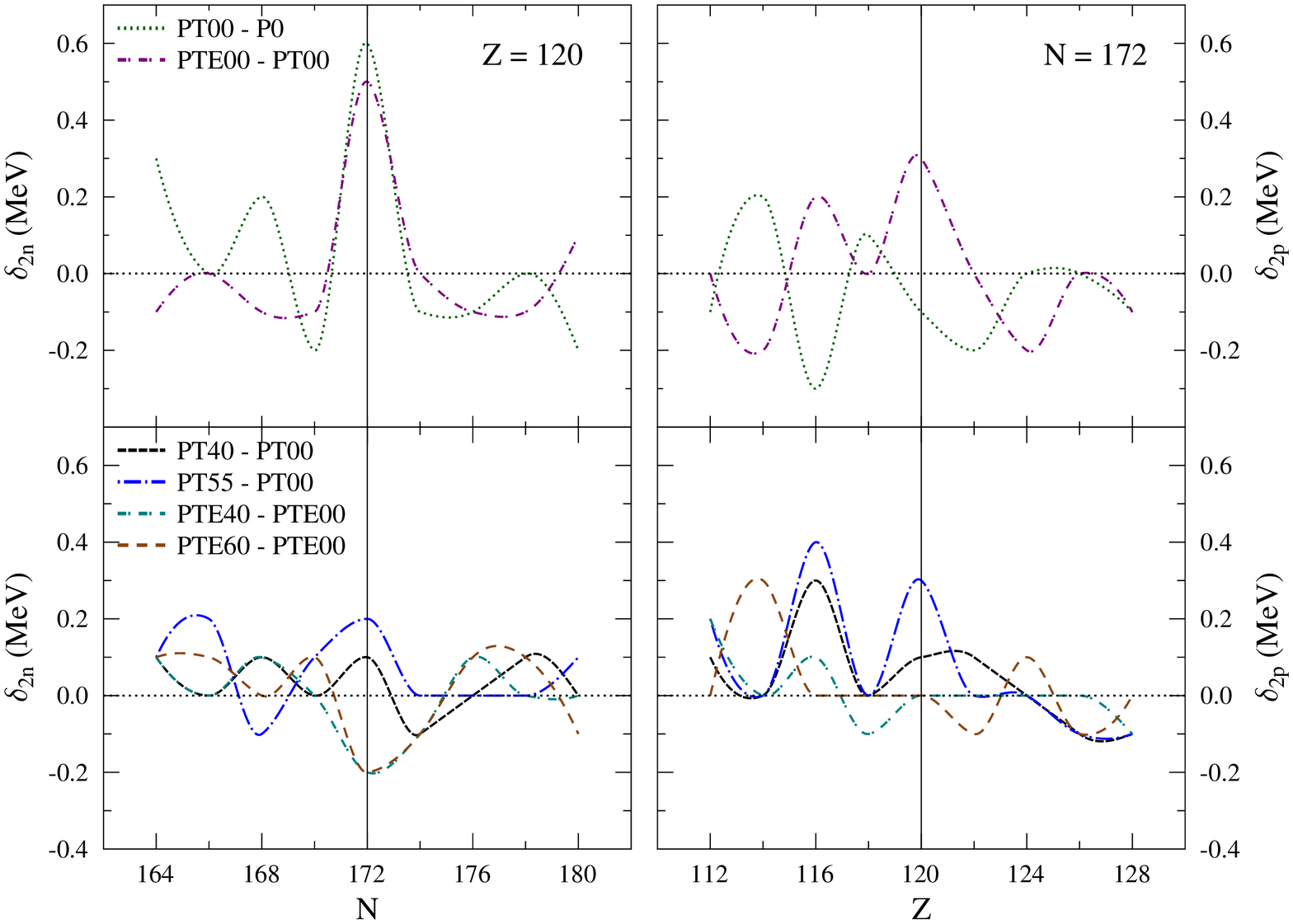, height=13.0cm, width=15.5cm }
\caption{{\footnotesize (Color online). The same as in Fig.~\ref{Fig:SHPb1}  but for the chain of $Z$ = 120 isotopes and the $N$ = 172 isotones. } }
\label{Fig:SH1201}
\end{figure}

Fig.~\ref{Fig:SH120}, shows the same observables as Fig.~\ref{Fig:SHPb} for  the chains of $Z$ = 120 isotopes and $N$ = 172 isotones.  The similar  role of these corresponding terms in the trend of binding energies and  two nucleons gaps to the  ones of  $Z$ = 82 isotopes and  $N$ = 126 isotones can be observed in  $Z$ = 120 isotopes and $N$ = 172 isotones. It can be observed in the corresponding plots that significant effects of tensor couplings  appear in the binding energies of $N$ $\le$ 185 and  $Z$ $\ge$ 115. In the  lower panel of the corresponding figure, it is obvious that the largest peaks of $\delta_{2p}$ and $\delta_{2n}$ predicted by all parameter sets used appear clear at $N$ = 172 for $Z$ = 120 isotopes and at $Z$ = 120 for $N$ = 172 isotones. However, it is interesting to observe  that due to the presence of tensor couplings and Coulomb exchange term, the magnitude of the largest peak of the gap at $N$ = 172 and also those at $N$ = 138 significantly increase. In this work, if tensor couplings are included, we do  not obtain strong evidence that $^{304}$120 is double magic nucleus because the peaks of two neutron gap at $N$ = 184 are relatively smaller compared to that of P0 parameter set.  Unless we obtain visible peak of neutron gap at $^{318}$120 in our calculation due to the role of tensor couplings, in general, the results of our two nucleons gaps  for $Z$ = 120 isotopes and  $N$ = 172 isotones are quite consistent to those obtained in Ref.~\cite{Jiang2010}. Similar to those of Fig.~\ref{Fig:SHPb1}, the detailed role of these terms for $\delta_{2n}$ of nuclei around  $N$ = 172 and  $Z$ = 120 can be seen clearly in Fig.~\ref{Fig:SH1201}. It can be observed for $Z$ = 120 and $N$ = 172 the crucial role of interplay among tensor couplings, Coulomb exchange term and various isoscalar-isovector couplings for gaps formation. It is interesting to observe that for  $Z$ = 120 isotopes, for the case  $\Lambda$ = 0, the effect of tensor couplings and Coulomb exchange term tends to relatively decrease  the magnitude of the peak. However, for the case  of non zero $\Lambda$, the parameter sets with Coulomb exchange term tend to decrease but those without Coulomb exchange tend to relatively increase  the magnitude of the peak. On the other hand,  like what happen with $N$ = 126 isotones, for  $N$ = 172 isotones, for the case  $\Lambda$ = 0, the role of tensor couplings tends to relatively decrease  but the role of  Coulomb exchange term tends to relatively increase  the magnitude of the gap. On the other hand, the role of isoscalar-isovector coupling relatively decreases  the magnitude of the gap where the quantitative effect also depends on the presence of Coulomb exchange term.

\subsection{Single Particle Spectra}
\begin{figure}
\epsfig{figure=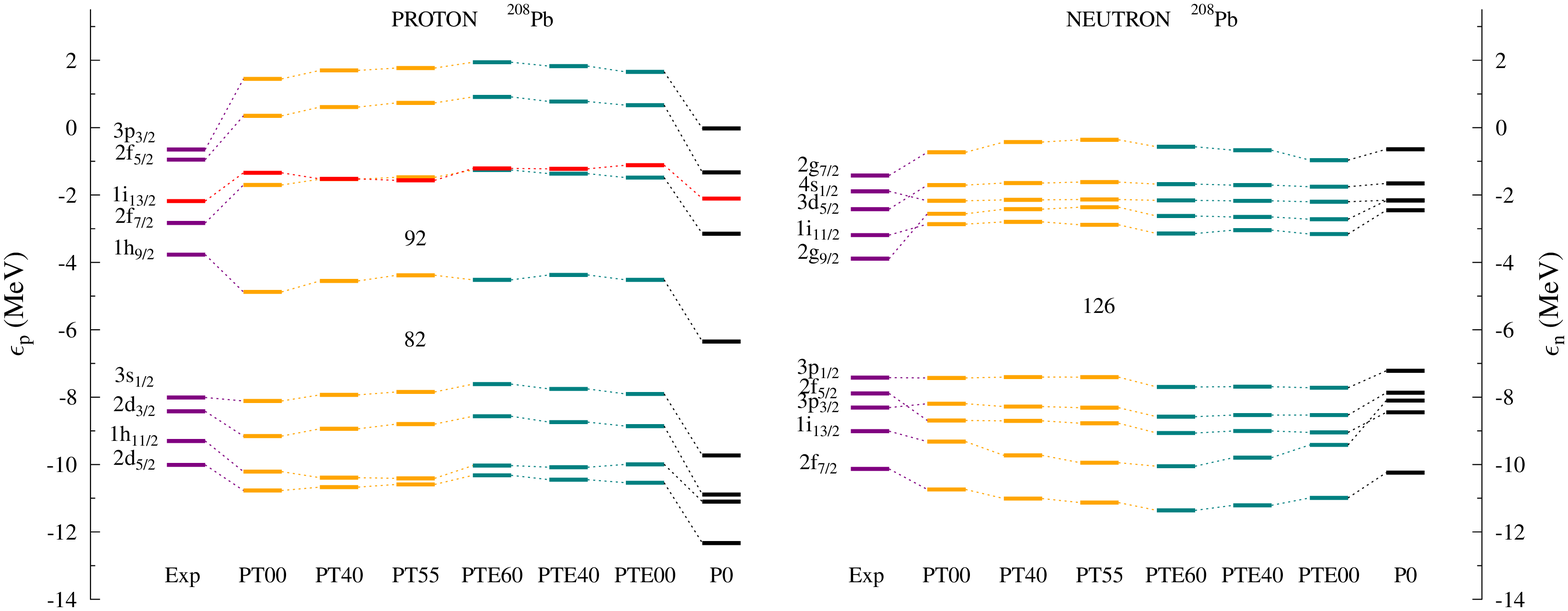, height=9cm, width=17cm }
\caption{{\footnotesize (Color online). Effects of tensor couplings and Coulomb exchange term with various isoscalar-isovector couplings  on SPSs of $^{208}$Pb.  }}
\label{Fig:Pbsps}
\end{figure}

Before we discuss the effects of inclusion tensor couplings, Coulomb exchange term and isoscalar-isovector couplings in RMF model on SPSs of  $^{292}$120 prediction, we will first discuss  the effects of the corresponding terms on proton ($\pi$) and neutron ($\nu$) SPS of  $^{208}$Pb. For comparison, we took SPSs experimental data  of $^{208}$Pb from Ref \cite{AMBh05}. It can be seen in Fig.~\ref{Fig:Pbsps} that all parameter sets used predict that the maximum gap  between $3s_{1/2}$ and $1h_{9/2}$ states in $\pi$ SPS appears for the $Z$ = 82, while the maximum gap between $3p_{1/2}$ and $1f_{11/2}$ states for $\nu$  SPS appears for $N$ = 126. Their prediction is in accordance with the experimental data. However, contrary to that from the experimental data, the corresponding parameter sets yield relatively large (spurious)  gaps  between $2f_{7/2}$ and $1h_{9/2}$ states in $\pi$  SPS for the $Z$ = 92. It is obvious that these  terms can not fix this spurious gap problem. It is reported in Refs.~\cite{LSGM2007,Meng2016X,Meng2016Z} that the appearance of the spurious shells at $Z$ = 58 ( $^{132}$Sn) and  $Z$ = 92 ($^{208}$Pb)  can be avoided only by including the exchange contribution of $\sigma$, $\omega$, $\rho$,  $\pi$, and $\rho$-tensor. It is important to note the appearance of this gap is generic in RMF  frameworks. Therefore, this artificial gap will  also appear in SHN. It can be observed in Fig.~\ref{Fig:Pbsps} also that the tensor couplings and Coulomb exchange term contributions in each state yield additional repulsive contribution to proton single particle energy. Therefore, each energy state in  $\pi$ SPS of  $^{208}$Pb will be shifted upward when these contributions are included. While for $\nu$ SPS of  $^{208}$Pb, the tensor couplings and Coulomb exchange term contributions yield additional attractive contribution to each energy state so that each energy state in $\nu$ SPS of $^{208}$Pb will be shifted downward when these contributions are included. In general, SPSs shifted due to the role of tensor couplings contribution is more pronounced than that of isoscalar-isovector coupling and Coulomb exchange term contributions. The presence of tensor couplings also makes the energy shifted under Fermi energy  of $\pi$ and $\nu$ SPSs of $^{208}$Pb more close to the experimental data. In general, the shifted energy magnitude due to these contributions in each state is state dependent. Furthermore, the effect of isoscalar-isovector coupling term in $\pi$ and $\nu$ SPSs is rather varied. In some states, when  we increase the $\Lambda$ value, the corresponding states are pushed upward but in some other states, they are pushed downward. However, in some states, we can also observe that there is indeed a strong correlation between the corresponding single particle energy and the  isoscalar-isovector coupling value. It is also interesting to observe that there is one state i.e. $\nu 4s_{1/2}$ which  is independent to  the variations of tensor couplings, isoscalar-isovector coupling and Coulomb exchange term contributions. 

\begin{figure}
\epsfig{figure=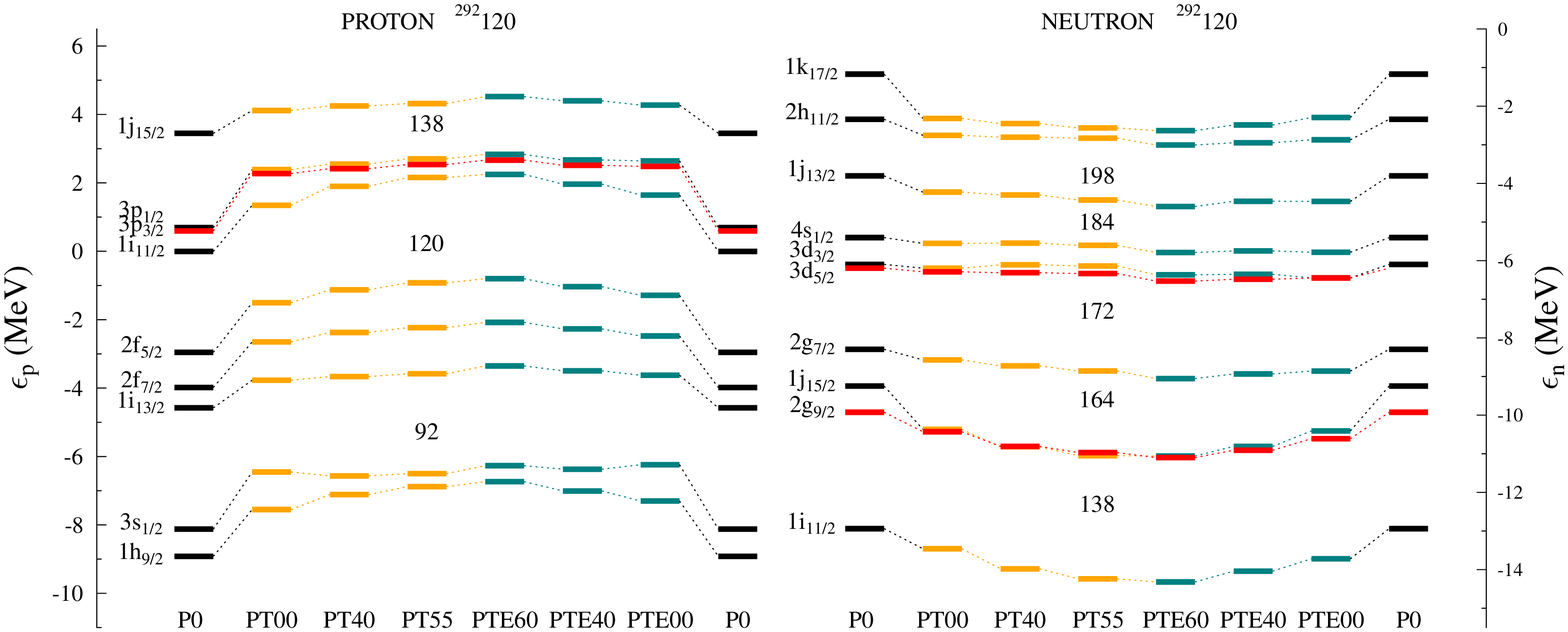, height=9cm, width=17cm }
\caption{{\footnotesize (Color online). The same as in Fig.~\ref{Fig:Pbsps} but for SPSs of $^{292}$120. }}
\label{Fig:SHsps}
\end{figure}

\begin{figure}
\epsfig{figure=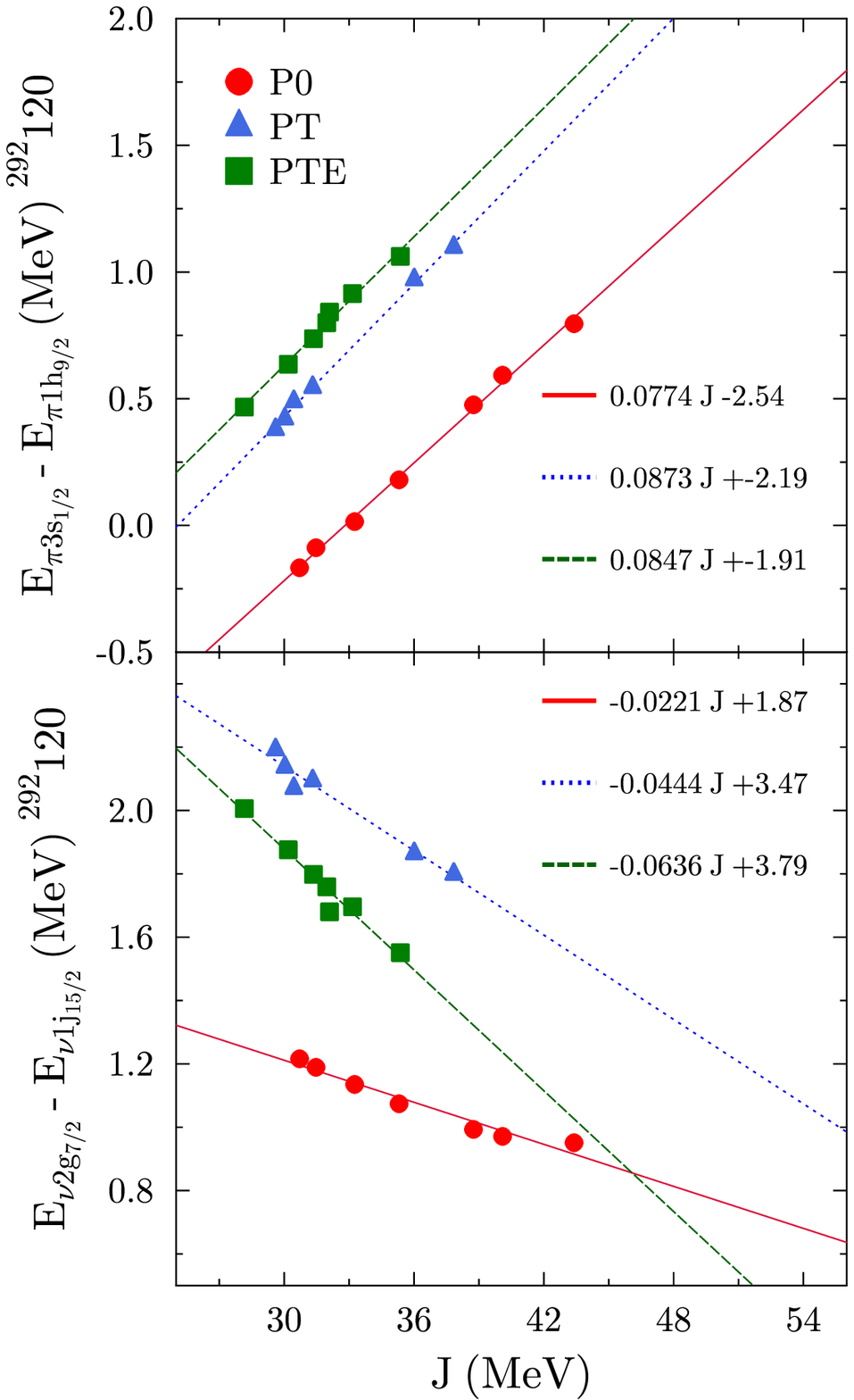, height=15cm, width=10.5cm }
\caption{{\footnotesize (Color online). Effects of tensor couplings and Coulomb exchange term with various isoscalar-isovector coupling values on the correlation of some SPSs with $J$ of $^{292}$120. Upper panel for proton while the lower panel for neutron SPSs. }}
\label{Fig:SHspsCorr}
\end{figure}

The effects of tensor couplings, isoscalar-isovector coupling and Coulomb exchange term contributions on the $\pi$ and $\nu$ SPSs of $^{292}$120 can be observed in Fig.~\ref{Fig:SHsps}, while the linear correlation of some SPS gaps of $^{292}$120 with $J$ is shown in Fig.~\ref{Fig:SHspsCorr}. It is obvious for $^{292}$120, the maximum gap in $\pi$ SPS appears in $Z$ = 120, while $\nu$  SPS the maximum gap is in $N$ = 172. These SPSs confirm that $^{292}$120 is double magic nucleus.  Qualitatively, the similar role of tensor couplings and Coulomb exchange term contributions to that occurring in $\nu$ and $\pi$ SPSs of $^{208}$Pb also appears in $\nu$ and $\pi$   SPSs of $^{292}$120. In general, if the skin thickness in $^{292}$120 is decreased by increasing $\Lambda$ then nearly all the corresponding SPSs are pushed upward. 
  
It can be seen from Fig.~\ref{Fig:SHspsCorr} that the gap  $\nu$ $E_{2g7/2}- E_{1j15/2}$ of $^{292}$120 decrease when $J$ increases while the gap $\pi$ $E_{3s1/2}- E_{1h9/2}$ of $^{292}$120 increase when $J$ increases. These linear correlations are consistent to those obtained by the author of Ref.~\cite{Jiang2010} for the case $\pi$ $E_{3s1/2}- E_{1h9/2}$ gap in $^{292}$120. It is also interesting to see that the effect of tensor couplings and Coulomb exchange contributions makes the magnitude of the corresponding gaps for the same $J$ different and they also make the slope of the linear relation formed in each corresponding gap different. 

Therefore, in general, the tensor couplings, isoscalar-isovector coupling, and Coulomb exchange term contributions significantly affect the spacing of single particle energies. Even, in couple states the ordering changes due to the roles of these terms. It means that these contributions can modify the position of each state of SHN and heavy nuclei. However, they  do not visibly influence the shell closures of SHN predictions. To this end, we should pay careful attention to the interpretation of the obtained SPSs  in this work because there are important effects which are not yet considered here. First the appearance of spurious gap can only be avoided  by introducing  exchange contributions of $\sigma$, $\omega$, $\rho$, $\pi$, and $\rho$-tensor. Second, it is known the SPS of  the corresponding nuclei are modified if empirical shift \cite{Afan06,Afan05}, deformation  and particle-vibration coupling were taken into account (see Refs.~\cite{BH2013,Vreten2002} and the references therein) in calculations. 

\begin{figure}
\epsfig{figure=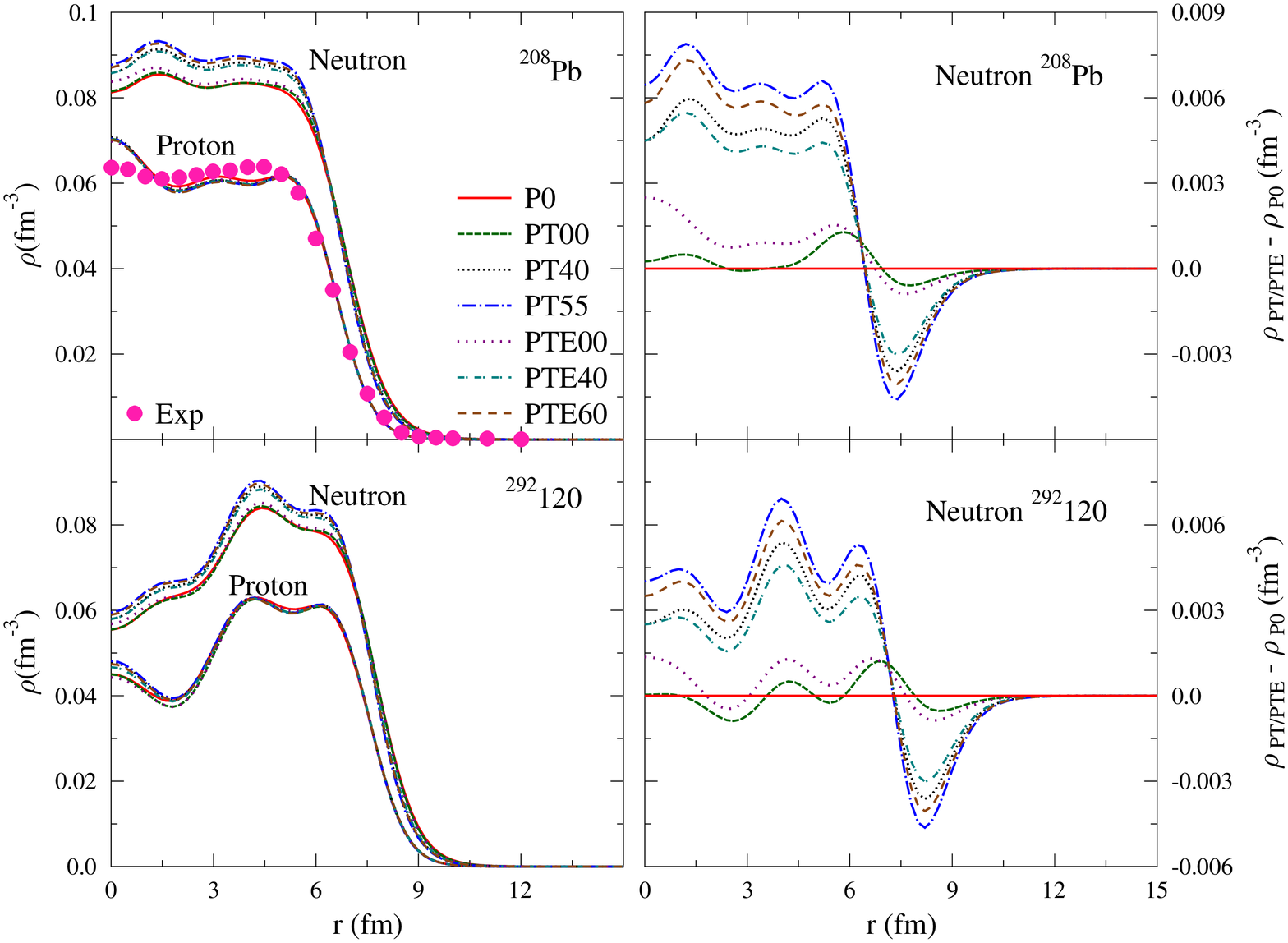, height=15cm, width=16.5cm }
\caption{{\footnotesize (Color online). Nucleon density distribution in the $^{208}$Pb (upper left panel) and $^{292}$120 (lower left panel) with various parameter sets. Effects of tensor couplings, Coulomb exchange term, and various isoscalar-isovector coupling values on the neutron density distribution of $^{208}$Pb (upper right panel) and $^{292}$120 (lower right panel).} }
\label{Fig:SHDens}
\end{figure}

\begin{figure}
\epsfig{figure=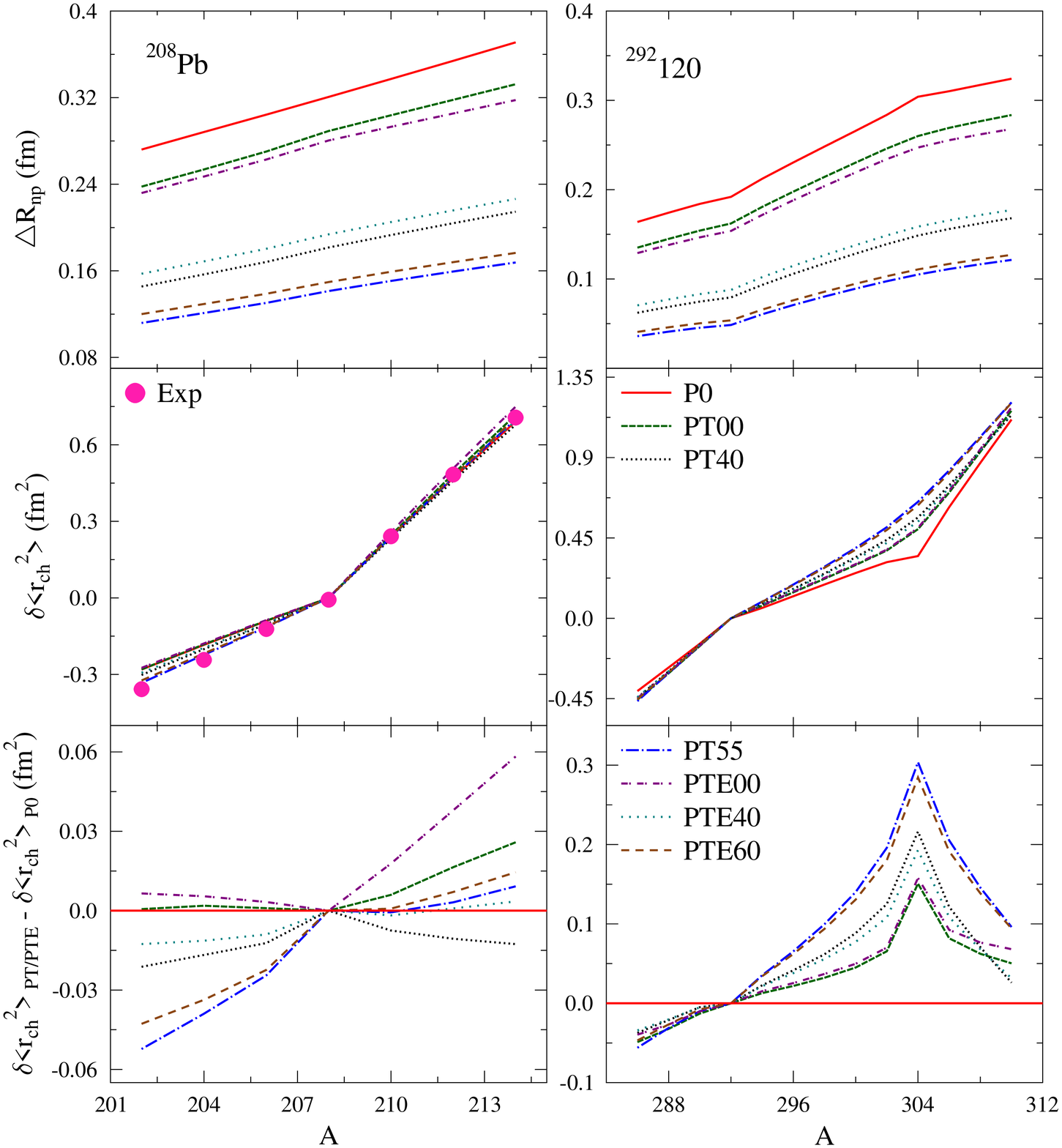, height=18.5cm, width=16.5cm }
\caption{{\footnotesize (Color online). Neutron skin thicknesses (upper panels) and mean square charge radius (middle panels) with various parameter sets for $Z$ = 82 (left panels) and $Z$ = 120 isotopes (right panels). Effects of tensor couplings, Coulomb exchange term, and various isoscalar-isovector coupling values on  the mean square charge radius (lower panels). The experimental data for the neutron skin thickness of $^{208}$Pb is obtained in Ref. \cite{warda} while for mean square charge radius is obtained in Ref. \cite{Goddard}.}}
\label{Fig:SHCSkn}
\end{figure}

\subsection{Density Profiles and Nuclei Radii}
For completeness, we have plotted the profiles of proton and neutron densities of  $^{208}$Pb and $^{292}$120 and the corresponding neutron densities differences in the left and right panels of Fig.~\ref{Fig:SHDens} as well as the skin thicknesses and mean squared charge radius of the $^{208}$Pb and $^{292}$120 isotopes in Fig.~\ref{Fig:SHCSkn}. The fact that the tensor couplings, isoscalar-isovector coupling and Coulomb exchange term contributions significantly affect the neutron densities can be seen more obvious in the right panels of Fig.~\ref{Fig:SHDens}. Furthermore, we need to point out that Coulomb exchange term contribution gives attractive effect when tensor couplings and isoscalar-isovector couplings are included. While the  isoscalar-isovector coupling is indeed the main factor in increasing neutron density distribution. The general central depression in density profiles is shown in the case $^{292}$120 for all parameter sets used. In the upper left panel of Fig. \ref{Fig:SHDens}, for comparison, we also show the experimental data of proton density of $^{208}$Pb, which are taken from Ref. \cite{Pei}. Except in the region close to the center, it can be observed that the proton density predicted by the used parameter sets is quite consistent  with the experimental data of proton density.

It is also interesting to observe in the upper panels of Fig.\ref{Fig:SHCSkn} that the prediction of neutron skin thicknesses ($\triangle{R_{np}}$) of $^{208}$Pb and $^{292}$120 nuclei is very sensitive to the predicted value of the corresponding nuclear matter symmetry energy at $\rho_0$. This fact is consistent with the finding of the authors of Refs.~\cite{Jiang2010} and~\cite{warda} where smaller symmetry energy at  $\rho_0$ will yield smaller $\triangle{R_{np}}$ values. Tensor couplings and  Coulomb exchange term  play a role in decreasing the  neutron skin thicknesses. It can be seen clearly that the Coulomb exchange term gives additional repulsive effect when isoscalar-isovector coupling is included. This is the reason why the $\triangle{R_{np}}$ values obtained by PTE60 and PTE40 (with Coulomb exchange term ) are higher than PT55 and PT40 (without Coulomb exchange term) , although PTE60 and PTE40 yield smaller symmetry energies than those of PT55 and PT40. If we look at $^{208}$Pb isotopes, $\triangle{R_{np}}$ versus $A$ relation which is predicted by all parameter sets is linear. While for $^{292}$120 isotopes, there is a kink appears along isotopes chain. Kink on neutron skin thicknesses along the isotope chains can also be used  to identify double magic nucleus \cite{warda}. Although in our calculation, no visible kink at $^{208}$Pb isotopes is clearly observed, the prediction of neutron skin thicknesses obtained by our parameter sets with tensor couplings and/or  Coulomb exchange term inclusion  is quite close to the experimental data, especially where $\Lambda$ = 0.06 (PTE60). In middle panels of Fig. \ref{Fig:SHCSkn}, we see the $\delta\left\langle{r_{ch}^{2}}\right\rangle$ i.e., the difference between the calculation of mean square charge radius  of each nucleus of $^{208}$Pb and $^{292}$120 isotopes and  $^{208}$Pb and $^{292}$120 nuclei, respectively. It can be observed that $\delta\left\langle{r_{ch}^{2}}\right\rangle$ of  $^{208}$Pb isotopes are not too sensitive to symmetry energy variation. However, the situation is rather different for the cases of  $^{292}$120 isotopes. It can also be seen that a kink appears in $A$ = 208 for $^{208}$Pb isotopes and $A$ = 292 and 304 for $^{292}$120 isotopes, respectively. And if we look at the result of $\delta\left\langle{r_{ch}^{2}}\right\rangle$ calculation  of $^{208}$Pb isotopes of each parameter set used, they are consistent to the one predicted by the experimental data. In lower panels of Fig. \ref{Fig:SHCSkn}, we plot the difference between $\delta\left\langle{r_{ch}^{2}}\right\rangle$ predicted by PTX (PTEX) and the one of PT0 to show more clearly the effects of these corresponding terms on  $\delta\left\langle{r_{ch}^{2}}\right\rangle$. It can be observed for  $^{208}$Pb isotopes, the corresponding terms give quite effect for nuclei of $^{208}$Pb isotopes with neutron significantly larger or smaller than N=126. Similar situation happens around A=292 for Z=120 isotopes. However, it is interesting to observe that around A=304 of Z=120 isotopes, all parameter sets used predict the kink appearance  in which the corresponding peak magnitude depends significantly on $\Lambda$ but depends only slightly on the role of Coulomb exchange term.    

\section{CONCLUSIONS}
\label{sec_conclu}
  
In conclusion, we have systematically investigated  the influence of tensor coupling terms, Coulomb exchange term and isoscalar-isovector coupling terms on nuclear matter, finite nuclei and SHN properties. For the case of nuclear matter,  we have shown that $J$, $L$, $K_{\rm asy}$ and $K_{\rm sat,2}$ decrease when the value of parameter $\Lambda$ increases.  We have  also found that for a fixed value of $\Lambda$, implicit inclusion of tensor coupling terms and Coulomb exchange term tends to slightly decrease  the  $K_0$, $J$, $L$, $K_{\rm asy}$ and $K_{\rm sat,2}$ value, respectively.  Furthermore, we found that the slope of the linear relation of neutron skin thickness of $^{208}$Pb versus $J$ does not significantly change  by the presence of tensor coupling terms and Coulomb exchange term. However, they change the constant value  of the corresponding linear relation. As the consequence, for the same fixed value of $J$ or $L$, the presence of  tensor coupling terms and Coulomb exchange term, respectively makes the neutron skin thickness of $^{208}$Pb predictions thicker. For the case of finite nuclei with quite wide range of mass spectrum, we have found that the roles of tensor coupling terms, Coulomb exchange term and various isoscalar-isovector coupling terms on nuclei bulk properties are not always the same for every nucleus but they are varied depending on the masses of the corresponding nuclei. On average,  the role of tensor coupling terms is quite significant for changing the prediction of finite nuclei bulk properties, while it seems for some nuclei, the experimental data of binding energies are less compatible with the presence of Coulomb exchange term in LDA and  they tend to disfavor the presence of isoscalar-isovector coupling term with too high $\Lambda$. For the case of SHE, we have found that tensor coupling terms, Coulomb exchange term and isoscalar-isovector coupling term influence detailed nuclei properties such as binding energies, the gaps magnitude of $\delta_{2p}$ and $\delta_{2n}$, SPSs, neutron densities, neutron skin thicknesses and mean squared charge radii. However, these terms can not affect the close shells predictions of double magic heavy ($^{208}$Pb)  and super-heavy ($^{292}$120) nuclei.

\section*{ACKNOWLEDGMENT}
A. S acknowledges support from the University of Indonesia  through Research Cluster Grant 2015 No. 1862/UN.R12/HKP.05.00/2015.

\newpage

\end{document}